# First-Order Stable Model Semantics and First-Order Loop Formulas

**Joohyung Lee** JOOLEE@ASU.EDU
**Yunsong Meng** YUNSONG.MENG@ASU.EDU
*School of Computing, Informatics,*
*and Decision Systems Engineering*
*Arizona State University*
*Tempe, AZ 85287, USA*

## Abstract

Lin and Zhao's theorem on loop formulas states that in the propositional case the stable model semantics of a logic program can be completely characterized by propositional loop formulas, but this result does not fully carry over to the first-order case. We investigate the precise relationship between the first-order stable model semantics and first-order loop formulas, and study conditions under which the former can be represented by the latter. In order to facilitate the comparison, we extend the definition of a first-order loop formula which was limited to a nondisjunctive program, to a disjunctive program and to an arbitrary first-order theory. Based on the studied relationship we extend the syntax of a logic program with explicit quantifiers, which allows us to do reasoning involving non-Herbrand stable models using first-order reasoners. Such programs can be viewed as a special class of first-order theories under the stable model semantics, which yields more succinct loop formulas than the general language due to their restricted syntax.

## 1. Introduction

According to the theorem on loop formulas (Lin & Zhao, 2004), the stable models of a logic program (Gelfond & Lifschitz, 1988) can be characterized as the models of the logic program that satisfy all its loop formulas. This idea has turned out to be widely applicable in relating the stable model semantics to propositional logic, and has resulted in an efficient method for computing answer sets using SAT solvers. Since the original invention of loop formulas for nondisjunctive logic programs by Lin and Zhao (2004), the theorem has been extended to more general classes of logic programs, such as disjunctive programs (Lee & Lifschitz, 2003), infinite programs and programs containing classical negation (Lee, 2005; Lee, Lierler, Lifschitz, & Yang, 2010), arbitrary propositional formulas under the stable model semantics (Ferraris, Lee, & Lifschitz, 2006), and programs containing aggregates (Liu & Truszczynski, 2006; You & Liu, 2008). The theorem has also been applied to other nonmonotonic formalisms, such as nonmonotonic causal theories (Lee, 2004) and McCarthy's circumscription (Lee & Lin, 2006). The notion of a loop was further refined as an "elementary loop" (Gebser & Schaub, 2005; Gebser, Lee, & Lierler, 2006, 2011). However, all this work is restricted to the propositional case. Variables contained in a program are first eliminated by grounding—the process which replaces every variable with every object constant—and then loop formulas are obtained from the ground program. As a result, loop formulas were defined as formulas in propositional logic.

Chen, Lin, Wang, and Zhang's definition (2006) of a first-order loop formula is different in that loop formulas are directly obtained from a non-ground program, so that they are first-order logic formulas which retain variables. However, since the semantics of a logic program that they refer to is based on grounding, these first-order loop formulas are simply understood as schemas for ground loop formulas, and only Herbrand models of the loop formulas were considered in this context.

The stable model semantics that does not involve grounding appeared a year later (Ferraris, Lee, & Lifschitz, 2007, 2011). The authors define the stable models of a first-order sentence $F$ as the models of the second-order sentence that is obtained by applying the "stable model operator" SM to $F$. The definition of SM is close to the definition of the circumscription operator CIRC (McCarthy, 1980, 1986). Under the first-order stable model semantics, logic programs are viewed as a special class of first-order theories. A similar definition of a stable model was given by Lin and Zhou (2011), via logic of knowledge and justified assumption (Lin & Shoham, 1992). The first-order stable model semantics is also closely related to quantified equilibrium logic (Pearce & Valverde, 2005), and indeed, Ferraris et al. (2011) showed that they are essentially equivalent.

A natural question arising is how first-order loop formulas and the first-order stable model semantics are related to each other. In general, the first-order stable model semantics is more expressive than first-order logic, and as such cannot be completely characterized by first-order loop formulas. Like circumscription, the concept of transitive closure can be represented in the first-order stable model semantics, but not in any set of first-order formulas, even if that set is allowed to be infinite.[1] However, as we show in this paper, understanding the precise relationship between them gives us insights into the first-order stable model semantics and its computational properties.

In order to facilitate the comparison, we extend the definition of a first-order loop formula which was limited to nondisjunctive programs, to disjunctive programs and to arbitrary first-order theories. Also we present a reformulation of SM$[F]$ in the style of loop formulas, which includes the characterization of a loop by a syntactic formula. From this formulation, we derive several conditions, under which a first-order theory under the stable model semantics can be equivalently rewritten as first-order loop formulas.

Based on the relationship between the first-order stable model semantics and first-order loop formulas, we extend the syntax of logic programs with explicit quantifiers, which may be useful in overcoming some limitations of traditional answer set programs in reasoning about non-Herbrand models. We define the semantics of such extended programs by identifying them as a special class of first-order theories under the stable model semantics. Such programs inherit from the general language the ability to handle nonmonotonic reasoning under the stable model semantics even in the absence of the unique name and the domain closure assumptions that are built into the grounding-based answer set semantics. On the other hand, the restricted syntax of an extended program leads to more succinct loop formulas. The following program $\Pi_1$ is a simple insurance policy example represented in

1. Vladimir Lifschitz, personal communication.

this syntax.

$$\begin{array}{rcl} \mathit{HasWife}(x) & \leftarrow & \exists y\ \mathit{Spouse}(x,y) \\ \mathit{HasWife}(x) & \leftarrow & \mathit{Man}(x),\ \mathit{Married}(x) \\ \mathit{Married}(x) & \leftarrow & \mathit{Man}(x),\ \mathit{HasWife}(x) \\ \exists w\ \mathit{Discount}(x,w) & \leftarrow & \mathit{Married}(x),\ \mathit{not}\ \exists z\ \mathit{Accident}(x,z). \end{array}$$

The second and the third rules express that $\mathit{Married}(x)$ and $\mathit{HasWife}(x)$ are synonymous to each other when $x$ is a $\mathit{Man}$. The last rule states that $x$ is eligible for some discount plan (with the name unknown) if $x$ is married and has no record of accident. The quantifier in the first rule can be dropped without affecting the meaning, but the other quantifiers cannot. We will say that a program $\Pi$ *entails* a query $F$ (under the stable model semantics) if every stable model of $\Pi$ satisfies $F$. For example,

- $\Pi_1$ conjoined with $\Pi_2 = \{\mathit{Man}(\mathit{John})\}$ entails each of $\neg\exists x\ \mathit{Married}(x)$ and $\neg\exists xy\ \mathit{Discount}(x,y)$.
- $\Pi_1 \cup \Pi_2$ conjoined with $\Pi_3 = \{\exists y\ \mathit{Spouse}(\mathit{John},y)\}$ entails neither $\neg\exists x\ \mathit{Married}(x)$ nor $\neg\exists xy\ \mathit{Discount}(x,y)$, but entails each of $\exists x\ \mathit{Married}(x)$, $\exists xy \mathit{Discount}(x,y)$, and $\forall xy(\mathit{Discount}(x,y) \rightarrow x = \mathit{John})$.
- $\Pi_1 \cup \Pi_2 \cup \Pi_3$ conjoined with $\Pi_4 = \{\exists z\ \mathit{Accident}(\mathit{John},z)\}$ does not entail $\forall xy(\mathit{Discount}(x,y) \rightarrow x = \mathit{John})$, but entails $\neg\exists w\ \mathit{Discount}(\mathit{John},w)$.

The nonmonotonic reasoning of this kind requires non-Herbrand models since the names (or identifiers) of discount plans, spouses and accident records may be unknown. However, the traditional answer set semantics is limited to Herbrand models due to the reference to grounding. By turning the program into first-order loop formulas we can automate the example reasoning using a first-order theorem prover.

The paper is organized as follows. The next section reviews the first-order stable model semantics by Ferraris et al. (2007, 2011). Section 3 reviews the theorem on first-order loop formulas by Chen et al. (2006) and extends it to disjunctive programs and to arbitrary first-order sentences, limiting attention to Herbrand stable models. Section 4 extends these results to allow non-Herbrand stable models as well (possibly allowing functions) under a certain semantic condition, and compare the first-order stable model semantics with loop formulas by reformulating the former in terms of the latter. In Section 5, we present a series of syntactic conditions that imply the semantic condition in Section 4. Section 6 provides an extension of logic programs that contain explicit quantifiers and shows how query answering for such extended programs can sometimes be reduced to entailment checking in first-order logic via loop formulas. In Section 7, the results are further extended to distinguish between intensional and non-intensional predicates. Related work is described in Section 8, and long proofs are given in Appendix A.

This article is an extended version of a conference paper by Lee and Meng (2008).

## 2. Review of the First-Order Stable Model Semantics

This review follows a journal paper by Ferraris et al. (2011) that extends a conference paper by the same authors (Ferraris et al., 2007) by distinguishing between intensional and non-intensional predicates.

A *formula* is defined the same as in first-order logic. A *signature* consists of *function constants* and *predicate constants*. Function constants of arity 0 are called *object constants*. We assume the following set of primitive propositional connectives and quantifiers:

$$\bot \text{ (falsity)},\ \wedge,\ \vee,\ \rightarrow,\ \forall,\ \exists\ .$$

$\neg F$ is an abbreviation of $F \rightarrow \bot$, symbol $\top$ stands for $\bot \rightarrow \bot$, and $F \leftrightarrow G$ stands for $(F \rightarrow G) \wedge (G \rightarrow F)$. We distinguish between atoms and atomic formulas as follows: an *atom* of a signature $\sigma$ is an $n$-ary predicate constant followed by a list of $n$ terms that can be formed from function constants in $\sigma$ (including object constants) and object variables; *atomic formulas* of $\sigma$ are atoms of $\sigma$, equalities between terms of $\sigma$, and the 0-place connective $\bot$.

The stable models of $F$ relative to a list of predicates $\mathbf{p} = (p_1, \dots, p_n)$ are defined via the *stable model operator with the intensional predicates* $\mathbf{p}$, denoted by $\text{SM}[F; \mathbf{p}]$.[2] Let $\mathbf{u}$ be a list of distinct predicate variables $u_1, \dots, u_n$ of the same length as $\mathbf{p}$. By $\mathbf{u} = \mathbf{p}$ we denote the conjunction of the formulas $\forall \mathbf{x}(u_i(\mathbf{x}) \leftrightarrow p_i(\mathbf{x}))$, where $\mathbf{x}$ is a list of distinct object variables of the same length as the arity of $p_i$, for all $i = 1, \dots, n$. By $\mathbf{u} \leq \mathbf{p}$ we denote the conjunction of the formulas $\forall \mathbf{x}(u_i(\mathbf{x}) \rightarrow p_i(\mathbf{x}))$ for all $i = 1, \dots, n$, and $\mathbf{u} < \mathbf{p}$ stands for $(\mathbf{u} \leq \mathbf{p}) \wedge \neg(\mathbf{u} = \mathbf{p})$. For any first-order sentence $F$, the expression $\text{SM}[F; \mathbf{p}]$ stands for the second-order sentence

$$F \wedge \neg \exists \mathbf{u}((\mathbf{u} < \mathbf{p}) \wedge F^*(\mathbf{u})), \tag{1}$$

where $F^*(\mathbf{u})$ is defined recursively:

- $p_i(\mathbf{t})^* = u_i(\mathbf{t})$ for any list $\mathbf{t}$ of terms;
- $F^* = F$ for any atomic formula $F$ (including $\bot$ and equality) that does not contain members of $\mathbf{p}$;
- $(F \wedge G)^* = F^* \wedge G^*$;
- $(F \vee G)^* = F^* \vee G^*$;
- $(F \rightarrow G)^* = (F^* \rightarrow G^*) \wedge (F \rightarrow G)$;
- $(\forall x F)^* = \forall x F^*$;
- $(\exists x F)^* = \exists x F^*$.

(There is no clause for negation here, because we treat $\neg F$ as shorthand for $F \rightarrow \bot$.)

A model of a sentence $F$ (in the sense of first-order logic) is called $\mathbf{p}$-*stable* if it satisfies $\text{SM}[F; \mathbf{p}]$. We will often simply write $\text{SM}[F]$ instead of $\text{SM}[F; \mathbf{p}]$ when $\mathbf{p}$ is the list of all predicate constants occurring in $F$, and call a model of $\text{SM}[F]$ simply a *stable* model of $F$. We distinguish between the terms "stable models" and "answer sets" as follows.[3] By $\sigma(F)$ we denote the signature consisting of the function and predicate constants occurring in $F$.

2. The intensional predicates $\mathbf{p}$ are the predicates that we "intend to characterize" by $F$.
3. The distinction is useful because in the first-order setting, stable models are no longer Herbrand interpretations and may not be represented by *sets* of atoms.

If $F$ contains at least one object constant, an Herbrand interpretation[4] of $\sigma(F)$ that satisfies $\text{SM}[F]$ is called an *answer set* of $F$. The answer sets of a logic program $\Pi$ are defined as the answer sets of the *FOL-representation* of $\Pi$ (i.e., the conjunction of the universal closures of implications corresponding to the rules).

**Example 1** *For program $\Pi$ that contains three rules*

$$\begin{array}{l} p(a) \\ q(b) \\ r(x) \leftarrow p(x), not\ q(x) \end{array}$$

*the FOL-representation $F$ of $\Pi$ is*

$$p(a) \wedge q(b) \wedge \forall x((p(x) \wedge \neg q(x)) \rightarrow r(x)) \tag{2}$$

*and* $\text{SM}[F]$ *is*

$$\begin{array}{l} p(a) \wedge q(b) \wedge \forall x((p(x) \wedge \neg q(x)) \rightarrow r(x)) \\ \quad \wedge \neg \exists uvw(((u,v,w) < (p,q,r)) \wedge u(a) \wedge v(b) \\ \qquad\qquad \wedge \forall x(((u(x) \wedge (\neg v(x) \wedge \neg q(x))) \rightarrow w(x)) \wedge ((p(x) \wedge \neg q(x)) \rightarrow r(x)))), \end{array}$$

*which is equivalent to the first-order sentence*

$$\forall x(p(x) \leftrightarrow x = a) \wedge \forall x(q(x) \leftrightarrow x = b) \wedge \forall x(r(x) \leftrightarrow (p(x) \wedge \neg q(x))) \tag{3}$$

*(See Example 3 in the work of Ferraris et al., 2007). The stable models of $F$ are any first-order models of (3). The only answer set of $F$ is the Herbrand model $\{p(a),\ q(b),\ r(a)\}$.*

## 3. First-Order Loop Formulas and Herbrand Models

We review the definition of a first-order loop formula for a nondisjunctive program given by Chen et al. (2006) and extend it to a disjunctive program and to an arbitrary first-order sentence.

### 3.1 Review of First-Order Loop Formulas Defined by Chen et al. (2006)

We call a formula *negative* if every occurrence of every predicate constant in it belongs to the antecedent of an implication. For instance, any formula of the form $\neg F$ is negative because this expression is shorthand for $F \rightarrow \bot$. An equality $t_1 = t_2$ is also negative because it contains no predicate constants.

A *nondisjunctive* program is a finite set of rules of the form

$$A \leftarrow B, N, \tag{4}$$

4. Recall that an *Herbrand interpretation* of a signature $\sigma$ (containing at least one object constant) is an interpretation of $\sigma$ such that its universe is the set of all ground terms of $\sigma$, and every ground term represents itself. An Herbrand interpretation can be identified with the set of ground atoms to which it assigns the value *true*.

where $A$ is an atom, $B$ is a set of atoms, and $N$ is a negative formula. The rules may contain function constants of positive arity.[5]

We will say that a nondisjunctive program is in *normal form* if, for all rules (4) in it, $A$ is of the form $p(\mathbf{x})$ where $\mathbf{x}$ is a list of distinct variables. It is clear that every program can be turned into normal form using equality in the body. For instance, $p(a, b) \leftarrow q(a)$ can be rewritten as $p(x, y) \leftarrow x{=}a,\ y{=}b,\ q(a)$.

Let $\Pi$ be a nondisjunctive program and let $Norm(\Pi)$ be a normal form of $\Pi$. By $\sigma(\Pi)$ we denote the signature consisting of function and predicate constants occurring in $\Pi$. Given a finite set $Y$ of atoms, we assume that $Norm(\Pi)$ does not contain variables in $Y$, by renaming the variables in $Norm(\Pi)$. The *(first-order) external support formula* of $Y$ for $\Pi$, denoted by $ES_\Pi(Y)$, is the disjunction of

$$\bigvee_{\theta : A\theta \in Y} \exists \mathbf{z} \Big( B\theta \wedge N\theta \wedge \bigwedge_{\substack{p(\mathbf{t}) \in B\theta \\ p(\mathbf{t}') \in Y}} (\mathbf{t} \neq \mathbf{t}') \Big) \tag{5}$$

for all rules (4) in $Norm(\Pi)$,[6] where $\theta$ is a substitution that maps variables in $A$ to terms occurring in $Y$, and $\mathbf{z}$ is the list of all variables that occur in

$$A\theta \leftarrow B\theta, N\theta$$

but not in $Y$.

The *(first-order) loop formula* of $Y$ for $\Pi$, denoted by $LF_\Pi(Y)$, is the universal closure of

$$\bigwedge Y \rightarrow ES_\Pi(Y). \tag{6}$$

(The expression $\bigwedge Y$ in the antecedent stands for the conjunction of all elements of $Y$.) When $\Pi$ is a propositional program, $LF_\Pi(Y)$ is equivalent to a conjunctive loop formula as defined by Ferraris et al. (2006).

The definition of a first-order dependency graph and the definition of a first-order loop are as follows. We say that an atom $p(\mathbf{t})$ *depends on* an atom $q(\mathbf{t}')$ in a rule (4) if $p(\mathbf{t})$ is $A$ and $q(\mathbf{t}')$ is in $B$. The *(first-order) dependency graph* of $\Pi$ is an infinite directed graph $(V, E)$ such that

- $V$ is the set of atoms of signature $\sigma(\Pi)$;[7]
- $(p(\mathbf{t})\theta, q(\mathbf{t}')\theta)$ is in $E$ if $p(\mathbf{t})$ depends on $q(\mathbf{t}')$ in a rule of $\Pi$ and $\theta$ is a substitution that maps variables in $\mathbf{t}$ and $\mathbf{t}'$ to terms (including variables) of $\sigma(\Pi)$.

A nonempty subset $L$ of $V$ is called a *(first-order) loop* of $\Pi$ if the subgraph of the first-order dependency graph of $\Pi$ induced by $L$ is strongly connected.

5. The original definition by Chen et al. (2006) does not allow function constants of positive arity.
6. For any lists of terms $\mathbf{t} = (t_1, \dots, t_n)$ and $\mathbf{t}' = (t'_1, \dots, t'_n)$ of the same length, $\mathbf{t} = \mathbf{t}'$ stands for $(t_1 = t'_1) \wedge \cdots \wedge (t_n = t'_n)$.
7. Note that $V$ is infinite since infinitely many object variables can be used to form atoms.

**Example 2** *Let* $\Pi$ *be the following program:*

$$\begin{array}{l} p(x) \leftarrow q(x) \\ q(y) \leftarrow p(y) \\ p(z) \leftarrow not\ r(z). \end{array} \tag{7}$$

*The following sets of atoms are first-order loops (among many others):* $Y_1 = \{p(u)\}$, $Y_2 = \{q(u)\}$, $Y_3 = \{r(u)\}$, $Y_4 = \{p(u), q(u)\}$. *Their loop formulas are*

$$\begin{array}{lcl} LF_\Pi(Y_1) &=& \forall u(p(u) \rightarrow (q(u) \lor \neg r(u))), \\ LF_\Pi(Y_2) &=& \forall u(q(u) \rightarrow p(u)), \\ LF_\Pi(Y_3) &=& \forall u(r(u) \rightarrow \bot), \\ LF_\Pi(Y_4) &=& \forall u(p(u) \land q(u) \rightarrow (q(u) \land u \neq u) \lor (p(u) \land u \neq u) \lor \neg r(u)). \end{array}$$

**Example 3** *Let* $\Pi$ *be the one-rule program*

$$p(x) \leftarrow p(y). \tag{8}$$

*Its finite first-order loops are* $Y_k = \{p(x_1), \dots, p(x_k)\}$ *where* $k > 0$. *Formula* $LF_\Pi(Y_k)$ *is*

$$\forall x_1 \dots x_k \Big( p(x_1) \land \dots \land p(x_k) \rightarrow \exists y(p(y) \land (y \neq x_1) \land \dots \land (y \neq x_k)) \Big). \tag{9}$$

The following is a reformulation of Theorem 1 from the work of Chen et al. (2006).

**Theorem 1** *Let* $\Pi$ *be a nondisjunctive program that contains at least one object constant but no function constants of positive arity, and let* $I$ *be an Herbrand interpretation of* $\sigma(\Pi)$ *that satisfies* $\Pi$.[8] *The following conditions are equivalent to each other:*

*(a)* $I$ *is a stable model of* $\Pi$;

*(b) for every nonempty finite set* $Y$ *of atoms of* $\sigma(\Pi)$, $I$ *satisfies* $LF_\Pi(Y)$; [9]

*(c) for every finite first-order loop* $Y$ *of* $\Pi$, $I$ *satisfies* $LF_\Pi(Y)$.

The sets of first-order loop formulas considered in conditions (b) and (c) above have obvious redundancies. For instance, the loop formula of $\{p(x)\}$ is equivalent to the loop formula of $\{p(y)\}$; the loop formula of $\{p(x), p(y)\}$ entails the loop formula of $\{p(z)\}$. Following the definition by Chen et al. (2006), given two sets of atoms $Y_1$ and $Y_2$, we say that $Y_1$ *subsumes* $Y_2$ if there is a substitution $\theta$ that maps variables in $Y_1$ to terms so that $Y_1\theta = Y_2$.

**Proposition 1** *(Chen et al., 2006, Proposition 7) For any nondisjunctive program* $\Pi$ *and any loops* $Y_1$ *and* $Y_2$ *of* $\Pi$, *if* $Y_1$ *subsumes* $Y_2$, *then* $LF_\Pi(Y_1)$ *entails* $LF_\Pi(Y_2)$.

Therefore in condition (c) from Theorem 1, it is sufficient to consider a set $\Gamma$ of loops such that, for every loop $L$ of $\Pi$, there is a loop $L'$ in $\Gamma$ that subsumes $L$. Chen et al. (2006) called such $\Gamma$ a *complete* set of loops. In Example 2, set $\{Y_1, Y_2, Y_3, Y_4\}$ is a finite complete set of loops of program (7). Program (8) in Example 3 has no finite complete set of loops.

8. We say that $I$ satisfies $\Pi$ if $I$ satisfies the FOL-representation of $\Pi$.
9. Note that $Y$ may contain variables.

### 3.2 Extension to Disjunctive Programs

A *disjunctive* program is a finite set of rules of the form

$$A \leftarrow B, N, \tag{10}$$

where $A$ and $B$ are sets of atoms, and $N$ is a negative formula. Similar to a nondisjunctive program, we say that a disjunctive program is in *normal form* if, for all rules (10) in it, all atoms in $A$ are of the form $p(\mathbf{x})$ where $\mathbf{x}$ is a list of distinct variables.

Let $\Pi$ be a disjunctive program and let $Norm(\Pi)$ be a normal form of $\Pi$. Given a finite set $Y$ of atoms, we first rename variables in $Norm(\Pi)$ so that no variables in $Norm(\Pi)$ occur in $Y$. The *(first-order) external support formula* of $Y$ for $\Pi$, denoted by $ES_\Pi(Y)$, is the disjunction of

$$\bigvee_{\theta: A\theta \cap Y \neq \emptyset} \exists \mathbf{z}\Big(B\theta \wedge N\theta \wedge \bigwedge_{\substack{p(\mathbf{t}) \in B\theta \\ p(\mathbf{t}') \in Y}} (\mathbf{t} \neq \mathbf{t}') \wedge \neg\big( \bigvee_{p(\mathbf{t}) \in A\theta} \big(p(\mathbf{t}) \wedge \bigwedge_{p(\mathbf{t}') \in Y} \mathbf{t} \neq \mathbf{t}'\big)\big)\Big) \tag{11}$$

for all rules (10) in $Norm(\Pi)$, where $\theta$ is a substitution that maps variables in $A$ to terms occurring in $Y$ or to themselves, and $\mathbf{z}$ is the list of all variables that occur in

$$A\theta \leftarrow B\theta, N\theta$$

but not in $Y$. The *(first-order) loop formula* of $Y$ for $\Pi$, denoted by $LF_\Pi(Y)$, is the universal closure of

$$\bigwedge Y \rightarrow ES_\Pi(Y).$$

Clearly, (11) is equivalent to (5) when $\Pi$ is nondisjunctive. When $\Pi$ and $Y$ are propositional, $LF_\Pi(Y)$ is equivalent to the conjunctive loop formula for a disjunctive program as defined by Ferraris et al. (2006).

**Example 4** *Let $\Pi$ be the program*

$$p(x, y) \ ; \ p(y, z) \ \leftarrow \ q(x)$$

*and let $Y = \{p(u, v)\}$. Formula $LF_\Pi(Y)$ is the universal closure of*

$$\begin{aligned} p(u, v) \rightarrow \quad & \exists z(q(u) \wedge \neg(p(v, z) \wedge ((v, z) \neq (u, v)))) \\ & \vee \exists x(q(x) \wedge \neg(p(x, u) \wedge ((x, u) \neq (u, v)))). \end{aligned}$$

Similar to the nondisjunctive case, we say that $p(\mathbf{t})$ *depends on* $q(\mathbf{t}')$ in $\Pi$ if there is a rule (10) in $\Pi$ such that $p(\mathbf{t})$ is in $A$ and $q(\mathbf{t}')$ is in $B$. The definitions of a first-order dependency graph and a first-order loop are extended to disjunctive programs in a straightforward way. Using these extended notions, the following theorem extends Theorem 1 to a disjunctive program. It is also a generalization of the main theorem by Ferraris et al. (2006) which was restricted to a propositional disjunctive program.

**Theorem 1**[d] *Let $\Pi$ be a disjunctive program that contains at least one object constant but no function constants of positive arity, and let $I$ be an Herbrand interpretation of $\sigma(\Pi)$ that satisfies $\Pi$. The following conditions are equivalent to each other:*

*(a) $I$ is a stable model of $\Pi$;*

*(b) for every nonempty finite set $Y$ of atoms of $\sigma(\Pi)$, $I$ satisfies $LF_{\Pi}(Y)$;*

*(c) for every finite first-order loop $Y$ of $\Pi$, $I$ satisfies $LF_{\Pi}(Y)$.*

### 3.3 Extension to Arbitrary Sentences

In this section we extend the definition of a first-order loop formula to an arbitrary first-order sentence.

As with a propositional loop formula defined for an arbitrary propositional theory (Ferraris et al., 2006), it is convenient to introduce a formula whose *negation* is close to *ES*. We define formula $NES_F(Y)$ (*"Negation of (First-order) External Support Formula"*), where $F$ is a first-order formula and $Y$ is a finite set of atoms, as follows. As before we assume that no variables in $Y$ occur in $F$, by renaming variables.

- $NES_{p_i(\mathbf{t})}(Y) = p_i(\mathbf{t}) \wedge \bigwedge_{p_i(\mathbf{t}')\in Y} \mathbf{t} \neq \mathbf{t}'$;
- $NES_{t_1=t_2}(Y) = (t_1 = t_2)$;
- $NES_{\bot}(Y) = \bot$;
- $NES_{F\wedge G}(Y) = NES_F(Y) \wedge NES_G(Y)$;
- $NES_{F\vee G}(Y) = NES_F(Y) \vee NES_G(Y)$;
- $NES_{F\rightarrow G}(Y) = (NES_F(Y) \rightarrow NES_G(Y)) \wedge (F \rightarrow G)$;
- $NES_{\forall xG}(Y) = \forall x NES_G(Y)$;
- $NES_{\exists xG}(Y) = \exists x NES_G(Y)$.

The *(first-order) loop formula* of $Y$ for $F$, denoted by $LF_F(Y)$, is the universal closure of

$$\bigwedge Y \rightarrow \neg NES_F(Y). \tag{12}$$

Note that the definition of *NES* looks similar to the definition of $F^*$ given in Section 2. When $F$ and $Y$ are propositional, $LF_F(Y)$ is equivalent to a conjunctive loop formula for a propositional formula that is defined by Ferraris et al. (2006). The following lemma tells us that the definition of a loop formula in this section generalizes the definition of a loop formula for a disjunctive program in the previous section.

**Lemma 1** *Let $\Pi$ be a disjunctive program in normal form, $F$ an FOL-representation of $\Pi$, and $Y$ a finite set of atoms. Formula $NES_F(Y)$ is equivalent to $\neg ES_{\Pi}(Y)$ under the assumption $F$.*

In order to extend the first-order dependency graph to an arbitrary formula, we introduce a few notions. We say that an occurrence of a subformula $G$ in a formula $F$ is *positive* if the number of implications in $F$ containing that occurrence in the antecedent is even; it is *strictly positive* if that number is 0. A *rule* of a first-order formula $F$ is an implication that occurs strictly positively in $F$. We will say that a formula is *rectified* if it has no variables that are both bound and free, and if all quantifiers in the formula refer to different variables. Any formula can be easily rewritten into a rectified formula by renaming bound variables.

We say that an atom $p(\mathbf{t})$ *depends on* an atom $q(\mathbf{t}')$ in an implication $G \rightarrow H$ if

- $p(\mathbf{t})$ has a strictly positive occurrence in $H$, and
- $q(\mathbf{t}')$ has a positive occurrence in $G$ that does not belong to any negative subformula of $G$.[10]

The definition of a first-order dependency graph is extended to formulas as follows. The *(first-order) dependency graph* of a rectified formula $F$ is the infinite directed graph $(V, E)$ such that

- $V$ is the set of atoms of signature $\sigma(F)$;
- $(p(\mathbf{t})\theta, q(\mathbf{t}')\theta)$ is in $E$ if $p(\mathbf{t})$ depends on $q(\mathbf{t}')$ in a rule of $F$ and $\theta$ is a substitution that maps variables in $\mathbf{t}$ and $\mathbf{t}'$ to terms of $\sigma(F)$.

Note that the rectified formula assumption is required in order to distinguish between dependency graphs of formulas such as

$$\forall x(p(x) \rightarrow q(x))$$

and

$$\forall x\, p(x) \rightarrow \forall x\, q(x).$$

Once the definition of a dependency graph is given, a loop of a first-order formula is defined in the same way as with a disjunctive program. Theorem 1 can be extended to first-order sentences using these extended notions.

**Theorem 1**$^f$ *Let $F$ be a rectified sentence that contains at least one object constant but no function constants of positive arity, and let $I$ be an Herbrand interpretation of $\sigma(F)$ that satisfies $F$. The following conditions are equivalent to each other:*

*(a) $I$ is a stable model of $F$ (i.e., $I$ satisfies* $\mathrm{SM}[F]$*);*

*(b) for every nonempty finite set $Y$ of atoms of $\sigma(F)$, $I$ satisfies $LF_F(Y)$;*

*(c) for every finite first-order loop $Y$ of $F$, $I$ satisfies $LF_F(Y)$.*

**Example 2 (continued)** *Consider the FOL-representation $F$ of the program in Example 2, for which $\{Y_1, Y_2, Y_3, Y_4\}$ is a complete set of loops. Under the assumption $F$,*

10. Recall the definition of a negative formula in Section 3.1.

- $LF_F(Y_1)$ *is equivalent to the universal closure of*

$$p(u) \to \neg\Big(\forall x(q(x) \to p(x) \wedge x \neq u) \wedge \forall y(p(y) \wedge y \neq u \to q(y)) \\ \wedge \forall z(\neg r(z) \to p(z) \wedge z \neq u)\Big);$$

- $LF_F(Y_2)$ *is equivalent to the universal closure of*

$$q(u) \to \neg\Big(\forall x(q(x) \wedge x \neq u \to p(x)) \wedge \forall y(p(y) \to q(y) \wedge y \neq u)\Big);$$

- $LF_F(Y_3)$ *is equivalent to the universal closure of*

$$r(u) \to \bot;$$

- $LF_F(Y_4)$ *is equivalent to the universal closure of*

$$p(u) \wedge q(u) \to \neg\Big(\forall x(q(x) \wedge x \neq u \to p(x) \wedge x \neq u) \\ \wedge \forall y(p(y) \wedge y \neq u \to q(y) \wedge y \neq u) \wedge \forall z(\neg r(z) \to p(z) \wedge z \neq u)\Big).$$

Proposition 1 can be straightforwardly extended to arbitrary sentences even without restricting the attention to loops.

**Proposition 1** $^f$ *For any sentence $F$ and any nonempty finite sets of atoms $Y_1$ and $Y_2$ of $\sigma(F)$, if $Y_1$ subsumes $Y_2$, then $LF_F(Y_1)$ entails $LF_F(Y_2)$.*

**Proof**. Note that $LF_F(Y_1)$ is

$$\forall \mathbf{z}\big(\bigwedge Y_1 \to \neg NES_F(Y_1)\big), \tag{13}$$

where $\mathbf{z}$ is the set of all variables in $Y_1$. If $Y_1$ subsumes $Y_2$, by definition, there is a substitution $\theta$ from variables in $Y_1$ to terms in $Y_2$ such that $Y_1\theta = Y_2$. It is clear that (13) entails

$$\forall \mathbf{z}'\big(\bigwedge Y_1\theta \to \neg NES_F(Y_1\theta)\big), \tag{14}$$

where $\mathbf{z}'$ is the set of all variables in $Y_1\theta$. (14) is exactly $LF_F(Y_2)$. □

Theorem 2 from the work of Ferraris et al. (2006) is a special case of Theorem $1^f$ when $F$ is restricted to a propositional formula.

**Corollary 1** *(Ferraris et al., 2006, Thm. 2) For any propositional formula $F$, the following formulas are equivalent to each other under the assumption $F$.*

*(a)* SM[$F$];

*(b) the conjunction of $LF_F(Y)$ for all nonempty sets $Y$ of atoms occurring in $F$;*

*(c) the conjunction of $LF_F(Y)$ for all (ground) loops $Y$ of $F$.*

## 4. Comparing First-Order Stable Model Semantics and First-Order Loop Formulas

The theorems in the previous section were restricted to Herbrand stable models. This section extends the results to allow non-Herbrand stable models as well, and compare the idea of loop formulas with SM by reformulating the latter in the style of loop formulas.

### 4.1 Loop Formulas Relative to an Interpretation

Recall that Theorem 1 and its extensions do not allow function constants of positive arity and are limited to Herbrand models of the particular signature obtained from the given theory. Indeed, the statements become wrong if these conditions are dropped.

**Example 5** *The following program contains a unary function constant $f$.*

$$\begin{array}{l} p(a) \\ p(x) \leftarrow p(f(x)). \end{array}$$

*The loops of this program are all singleton sets of atoms, and their loop formulas are satisfied by the Herbrand model $\{p(a), p(f(a)), p(f(f(a))), \dots\}$ of the program, but this model is not stable.*

**Example 3 (continued)** *The mismatch can happen even in the absence of function constants of positive arity. Consider the program in Example 3 and an interpretation $I$ such that the universe is the set of all integers, and $p^I$ contains all integers. Interpretation $I$ satisfies all first-order loop formulas (9), but it is not a stable model.*

These examples suggest that the mismatch between the first-order stable model semantics and the first-order loop formulas is related to the presence of an infinite path in the dependency graph that visits infinitely many vertices. In the following we will make this idea more precise, and extend Theorem $1^f$ to allow non-Herbrand interpretations under a certain condition.

First, we define a dependency graph *relative to an interpretation*. Let $F$ be a rectified formula whose signature is $\sigma$ and let $I$ be an interpretation of $\sigma$. For each element $\xi$ of the universe $|I|$ of $I$, we introduce a new symbol $\xi^\diamond$, called an *object name*. By $\sigma^I$ we denote the signature obtained from $\sigma$ by adding all object names $\xi^\diamond$ as additional object constants. We will identify an interpretation $I$ of signature $\sigma$ with its extension to $\sigma^I$ defined by $I(\xi^\diamond) = \xi$ (For details, see the work of Lifschitz, Morgenstern, & Plaisted, 2008).

The *dependency graph of $F$ w.r.t. $I$* is the directed graph $(V, E)$ where

- $V$ is the set of all atoms of the form $p_i(\boldsymbol{\xi}^\diamond)$ where $p_i$ belongs to $\sigma(F)$ and $\boldsymbol{\xi}^\diamond$ is a list of object names for $|I|$, and
- $(p_i(\boldsymbol{\xi}^\diamond), p_j(\boldsymbol{\eta}^\diamond))$ is in $E$ if there are atoms $p_i(\mathbf{t})$, $p_j(\mathbf{t}')$ such that $p_i(\mathbf{t})$ depends on $p_j(\mathbf{t}')$ in a rule of $F$ and there is a substitution $\theta$ that maps variables in $\mathbf{t}$ and $\mathbf{t}'$ to object names such that $(\mathbf{t}\theta)^I = \boldsymbol{\xi}$ and $(\mathbf{t}'\theta)^I = \boldsymbol{\eta}$.

We call a nonempty subset $L$ of $V$ a *loop of $F$ w.r.t. $I$* if the subgraph of the dependency graph of $F$ w.r.t. $I$ that is induced by $L$ is strongly connected. We say that $F$ is *bounded w.r.t. $I$* if every infinite path in the dependency graph of $F$ w.r.t. $I$ whose vertices are satisfied by $I$ visits only finitely many vertices. If $F$ is bounded w.r.t. $I$, then, clearly, every loop $L$ of $F$ w.r.t. $I$ such that $I \models L$ is finite. The definition is extended to a non-rectified formula by first rewriting it as a rectified formula. It also applies to the program syntax by referring to its FOL-representation.

**Theorem 2** *Let $F$ be a rectified sentence of a signature $\sigma$ (possibly containing function constants of positive arity), and let $I$ be an interpretation of $\sigma$ that satisfies $F$. If $F$ is bounded w.r.t. $I$, then the following conditions are equivalent to each other:*

*(a) $I \models \text{SM}[F]$;*

*(b) for every nonempty finite set $Y$ of atoms formed from predicate constants in $\sigma(F)$ and object names for $|I|$, $I$ satisfies $LF_F(Y)$;*

*(c) for every finite loop $Y$ of $F$ w.r.t. $I$, $I$ satisfies $LF_F(Y)$.*

The condition that $F$ is bounded w.r.t. $I$ is sufficient for ensuring the equivalence among (a), (b), and (c), but it is not a necessary condition. For instance, consider $F$ to be

$$\forall x\, p(x) \wedge \forall xy(p(x) \rightarrow p(y))$$

and $I$ to be a model of $F$ whose universe is infinite. Formula $F$ is not bounded w.r.t. $I$, but $I$ satisfies every loop formula, as well as $\text{SM}[F]$.

When $I$ is an Herbrand model of $\sigma(F)$, the dependency graph of $F$ w.r.t. $I$ is isomorphic to the subgraph of the first-order dependency graph of $F$ that is induced by vertices containing ground atoms. A set of ground atoms of $\sigma(F)$ is a loop of $F$ iff it is a loop of $F$ w.r.t. $I$. Hence Theorem 2 is essentially a generalization of Theorem $1^f$.

Note that the programs considered in Examples 3 and 5 are not bounded w.r.t. the interpretations considered there.

Clearly, if the universe of $I$ is finite, then $F$ is bounded w.r.t. $I$. This fact leads to the following corollary.

**Corollary 2** *For any rectified sentence $F$ and any model $I$ of $F$ whose universe is finite, conditions (a), (b), and (c) of Theorem 2 are equivalent to each other.*

In view of Proposition $1^f$ and Corollary 2, if the size of the universe is known to be a finite number $n$, it is sufficient to consider at most $2^{|\mathbf{p}|} - 1$ loop formulas, where $\mathbf{p}$ is the set of all predicate constants occurring in the sentence. Each loop formula is to check the external support of $\bigcup_{p \in K}\{p(\mathbf{x}_1), \dots, p(\mathbf{x}_{n^r})\}$ for each $K$ where

- $K$ is a nonempty subset of $\mathbf{p}$;
- $r$ is the arity of $p$ and each $\mathbf{x}_i$ is a list of variables of the length $r$ such that all variables in $\mathbf{x}_1, \dots, \mathbf{x}_{n^r}$ are pairwise distinct.

For instance, consider program (8). If the size of the universe is known to be 3, it is sufficient to consider only one loop formula (9) where $k = 3$.

Theorem $1^f$ essentially follows from Corollary 2 as the Herbrand universe of $\sigma(F)$ is finite when $F$ contains no function constants of positive arity.

Another corollary to Theorem 2 is acquired when $F$ has only "trivial" loops. We say that a formula $F$ is *atomic-tight* w.r.t. $I$ if every path in the dependency graph of $F$ w.r.t. $I$ whose vertices are satisfied by $I$ is finite. Clearly, this is a special case of boundedness condition, and every loop $L$ of an atomic-tight formula $F$ w.r.t. $I$ such that $I \models L$ is a singleton. The following is a corollary to Theorem 2, which tells us the condition under which stable models can be characterized by loop formulas of singleton loops only. By SLF$[F]$ ("loop formulas of singletons") we denote

$$\begin{array}{l}\{LF_F(\{p(\mathbf{x})\}) \mid p \text{ is a predicate constant in } \sigma(F)\text{, and } \mathbf{x} \text{ is a list} \\ \qquad \text{of distinct object variables whose length is the same as the arity of } p\}.\end{array} \tag{15}$$

**Corollary 3** *Let $F$ be a rectified sentence (possibly containing function constants of positive arity), and let $I$ be a model of $F$. If $F$ is atomic-tight w.r.t. $I$, then $I$ satisfies* SM$[F]$ *iff $I$ satisfies* SLF$[F]$.

SLF$[F]$ is similar to Clark's completion. In the propositional case, the relationship between the loop formulas of singletons and the completion is studied by Lee (2005). Below we describe their relationship in the first-order case. A sentence $F$ is in *Clark normal form* (Ferraris et al., 2011) if it is a conjunction of formulas of the form

$$\forall \mathbf{x}(G \rightarrow p(\mathbf{x})), \tag{16}$$

one for each predicate constant $p$ occurring in $F$, where $\mathbf{x}$ is a list of distinct variables, and $G$ has no free variables other than $\mathbf{x}$. The *completion* of a sentence $F$ in Clark normal form, denoted by Comp$[F]$, is obtained from $F$ by replacing each conjunctive term (16) with

$$\forall \mathbf{x}(p(\mathbf{x}) \leftrightarrow G).$$

Any nondisjunctive program can be turned into Clark normal form (Ferraris et al., 2011, Section 6.1).

**Corollary 4** *Let $F$ be the FOL-representation of a nondisjunctive program $\Pi$, and let $F'$ be the Clark normal form of $F$ as obtained by the process described in the work of Ferraris et al. (2011, Section 6.1). If $F$ is atomic-tight w.r.t. an interpretation $I$, then $I \models$* SM$[F]$ *iff $I \models$* Comp$[F']$.

**Proof**. Since $F$ is atomic-tight w.r.t. $I$, by Corollary 3, $I \models$ SM$[F]$ iff $I \models F \wedge$ SLF$[F]$. It is sufficient to show that, for each predicate constant $p$ occurring in $F$, under the assumption that $F$ is atomic-tight w.r.t. $I$,

$$I \models \forall \mathbf{x}\Big(p(\mathbf{x}) \rightarrow \bigvee_{p(\mathbf{t}') \leftarrow B, N \in \Pi} \exists \mathbf{z}\big((\mathbf{x} = \mathbf{t}') \wedge B \wedge N \wedge \bigwedge_{p(\mathbf{t}) \in B} (\mathbf{t} \neq \mathbf{x})\big)\Big) \tag{17}$$

iff

$$I \models \forall \mathbf{x}\Big(p(\mathbf{x}) \rightarrow \bigvee_{p(\mathbf{t}')\leftarrow B,N\in\Pi} \exists \mathbf{z}\big((\mathbf{x}=\mathbf{t}') \wedge B \wedge N\big)\Big), \tag{18}$$

where $\mathbf{z}$ is the list of all free variables in $p(\mathbf{x}) \leftarrow (\mathbf{x}=\mathbf{t}'), B, N$ that are not in $\mathbf{x}$.

Note that (17) is equivalent to saying that

$$I \models \forall \mathbf{x}\Big(p(\mathbf{x}) \rightarrow \bigvee_{p(\mathbf{t}')\leftarrow B,N\in\Pi} \exists \mathbf{z}\big((\mathbf{x}=\mathbf{t}') \wedge B \wedge N \wedge \bigwedge_{p(\mathbf{t})\in B} (\mathbf{t}\neq\mathbf{t}')\big)\Big). \tag{19}$$

From the assumption that $F$ is atomic-tight w.r.t. $I$, it follows that, for any rule $p(\mathbf{t}') \leftarrow B, N$ in $\Pi$ and any atom of $p(\mathbf{t})$ in $B$, $I \models \forall \mathbf{y}(\mathbf{t} \neq \mathbf{t}')$, where $\mathbf{y}$ is the list of all variables in $\mathbf{t}$ and $\mathbf{t}'$ (otherwise we find a singleton loop with a self-cycle, which contradicts that $F$ is atomic tight w.r.t. $I$). Consequently, (19) is equivalent to (18). $\square$

For example, let $F$ be the FOL-representation of the program

$$\begin{array}{rcl} p(b) & \leftarrow & p(a) \\ & \leftarrow & a \neq b \end{array} \tag{20}$$

SLF$[p(a) \rightarrow p(b)]$ is $\forall x(p(x) \rightarrow x = b \wedge p(a) \wedge x \neq a)$, while Comp$[\forall x(x = b \wedge p(a) \rightarrow p(x))]$ is $\forall x(p(x) \leftrightarrow x = b \wedge p(a))$. The additional conjunctive term $x \neq a$ can be dropped when we consider any model $I$ of $F$, for which $a^I \neq b^I$.

Corollary 4 is an enhancement of Theorem 11 from the work of Ferraris et al. (2011), which states the equivalence between SM$[F]$ and Comp$[F]$ for any tight sentence $F$ in Clark normal form. (Tight sentences are defined in a similar way, but in terms of a predicate dependency graph, whose vertices are predicate constants instead of atoms.) Every tight sentence is atomic-tight w.r.t. any model of the sentence. On the other hand, program (20) is atomic-tight w.r.t. any model of the program, but is not tight.

Theorem 2 tells us that one of the limitations of first-order loop formulas is that, even if infinitely many first-order loop formulas are considered, they cannot ensure the external support of a certain infinite set that forms an infinite path in the dependency graph of $F$ w.r.t. $I$. In the next section, by reformulating SM$[F]$, we show that the definition of SM$[F]$ essentially encompasses loop formulas, ensuring the external support of any sets of atoms, including those "difficult" infinite sets.

### 4.2 A Reformulation of SM

As before, let $F$ be a first-order formula of a signature $\sigma$, let $\mathbf{p} = (p_1, \dots, p_n)$ be the list of all predicate constants occurring in $F$, and let $\mathbf{u}$ and $\mathbf{v}$ be lists of predicate variables of the same length as $\mathbf{p}$. We define $NSES_F(\mathbf{u})$ ("*Negation of Second-Order External Support Formula*") recursively as follows.

- $NSES_{p_i(\mathbf{t})}(\mathbf{u}) = p_i(\mathbf{t}) \wedge \neg u_i(\mathbf{t})$;
- $NSES_{t_1=t_2}(\mathbf{u}) = (t_1 = t_2)$;
- $NSES_{\bot}(\mathbf{u}) = \bot$;

- $NSES_{F\wedge G}(\mathbf{u}) = NSES_F(\mathbf{u}) \wedge NSES_G(\mathbf{u})$;
- $NSES_{F\vee G}(\mathbf{u}) = NSES_F(\mathbf{u}) \vee NSES_G(\mathbf{u})$;
- $NSES_{F\rightarrow G}(\mathbf{u}) = (NSES_F(\mathbf{u}) \rightarrow NSES_G(\mathbf{u})) \wedge (F\rightarrow G)$;
- $NSES_{\forall xF}(\mathbf{u}) = \forall x NSES_F(\mathbf{u})$;
- $NSES_{\exists xF}(\mathbf{u}) = \exists x NSES_F(\mathbf{u})$.

**Lemma 2** *Let $F$ be a rectified sentence of a signature $\sigma$, $I$ an interpretation of $\sigma$, $\mathbf{p}$ the list of predicate constants occurring in $F$, $\mathbf{q}$ a list of predicate names* [11] *of the same length as $\mathbf{p}$ and $Y$ a set of atoms formed from predicate constants from $\sigma(F)$ and object names such that*

$$p_i(\boldsymbol{\xi}^\diamond) \in Y \text{ iff } I \models q_i(\boldsymbol{\xi}^\diamond),$$

*where $\boldsymbol{\xi}^\diamond$ is a list of object names. If $Y$ is finite, then*

$$I \models NSES_F(\mathbf{q}) \quad \text{iff} \quad I \models NES_F(Y).$$

**Proof**. By induction on $F$. We only list the case when $F$ is an atom. The other cases are straightforward. Let $F$ be an atom $p_i(\boldsymbol{\xi}^\diamond)$.

$$\begin{array}{ll} & I \models NSES_F(\mathbf{q}) \\ \text{iff} & I \models p_i(\boldsymbol{\xi}^\diamond) \wedge \neg q_i(\boldsymbol{\xi}^\diamond) \\ \text{iff} & I \models p_i(\boldsymbol{\xi}^\diamond) \text{ and } p_i(\boldsymbol{\xi}^\diamond) \notin Y \\ \text{iff} & I \models p_i(\boldsymbol{\xi}^\diamond) \text{ and for all } \boldsymbol{\eta}^\diamond \text{ such that } p_i(\boldsymbol{\eta}^\diamond) \in Y, \text{ it holds that } \boldsymbol{\xi}^\diamond \neq \boldsymbol{\eta}^\diamond \\ \text{iff} & I \models p_i(\boldsymbol{\xi}^\diamond) \wedge \bigwedge_{p_i(\boldsymbol{\eta}^\diamond)\in Y} \boldsymbol{\xi}^\diamond \neq \boldsymbol{\eta}^\diamond \\ \text{iff} & I \models NES_F(Y). \end{array}$$

□

SM[$F$] can be written in terms of *NSES* as follows. By *Nonempty*($\mathbf{u}$) we denote the formula

$$\exists \mathbf{x}^1 u_1(\mathbf{x}^1) \vee \cdots \vee \exists \mathbf{x}^n u_n(\mathbf{x}^n),$$

where each $\mathbf{x}^i$ is a list of distinct variables whose length is the same as the arity of $p_i$.

**Proposition 2** *For any sentence $F$,* SM[$F$] *is equivalent to*

$$F \wedge \forall \mathbf{u}((\mathbf{u} \leq \mathbf{p}) \wedge Nonempty(\mathbf{u}) \rightarrow \neg NSES_F(\mathbf{u})). \quad (21)$$

Now we represent the notion of a loop by a second-order formula. Given a rectified formula $F$, by $E_F(\mathbf{v}, \mathbf{u})$ we denote

$$\bigvee_{\substack{(p_i(\mathbf{t}),p_j(\mathbf{t}')) : \\ p_i(\mathbf{t}) \text{ depends on } p_j(\mathbf{t}') \text{ in a rule of } F}} \exists \mathbf{z}(v_i(\mathbf{t}) \wedge u_j(\mathbf{t}') \wedge \neg v_j(\mathbf{t}')),$$

11. Like object names, for every $n > 0$, each subset of $|I|^n$ has a name, which is an $n$-ary predicate constant not from the underlying signature.

where $\mathbf{z}$ is the list of all object variables in $\mathbf{t}$ and $\mathbf{t}'$. By $Loop_F(\mathbf{u})$ we denote the second-order formula

$$Nonempty(\mathbf{u}) \wedge \forall \mathbf{v}((\mathbf{v} < \mathbf{u}) \wedge Nonempty(\mathbf{v}) \rightarrow E_F(\mathbf{v}, \mathbf{u})). \tag{22}$$

Formula (22) represents the concept of a loop without referring to the notion of a dependency graph explicitly. This is based on the following observation. Consider a finite propositional program $\Pi$. A nonempty set $U$ of atoms that occur in $\Pi$ is a loop of $\Pi$ iff, for every nonempty proper subset $V$ of $U$, there is an edge from an atom in $V$ to an atom in $U \setminus V$ in the dependency graph of $\Pi$ (Gebser et al., 2006).

Recall the definition of a dependency graph relative to an interpretation. Let $F$ be a rectified sentence of a signature $\sigma$, and let $I$ be an interpretation of $\sigma$. The following proposition describes the relationship between formula (22) and a loop of $F$ w.r.t. $I$.

**Proposition 3** *Let $\mathbf{q}$ be a list of predicate names corresponding to $\mathbf{p}$, and let $Y$ be a set of atoms in the dependency graph of $F$ w.r.t. $I$ such that*

$$p_i(\boldsymbol{\xi}^\diamond) \in Y \text{ iff } I \models q_i(\boldsymbol{\xi}^\diamond),$$

*where $\boldsymbol{\xi}^\diamond$ is a list of object names. Then $I \models Loop_F(\mathbf{q})$ iff $Y$ is a loop of $F$ w.r.t. $I$.*

One might expect that, similar to the equivalence between conditions (a) and (c) from Theorem 2, formula SM$[F]$ is equivalent to the following formula:

$$F \wedge \forall \mathbf{u}((\mathbf{u} \leq \mathbf{p}) \wedge Loop_F(\mathbf{u}) \rightarrow \neg NSES_F(\mathbf{u})). \tag{23}$$

However, the equivalence does not hold in general, as the following example illustrates.

**Example 6** *Consider the FOL-representation $F$ of the following program*

$$p(x, y) \leftarrow q(x, z)$$
$$q(x, z) \leftarrow p(y, z),$$

*and an interpretation $I$ whose universe is the set of all nonnegative integers such that*

$$p^I = \{(m, m) \mid m \text{ is a nonnegative integer}\},$$
$$q^I = \{(m, m{+}1) \mid m \text{ is a nonnegative integer}\}.$$

*Formula $F$ is not bounded w.r.t. $I$ since the dependency graph of $F$ w.r.t. $I$ contains an infinite path such as*

$$\langle p(0^\diamond, 0^\diamond), q(0^\diamond, 1^\diamond), p(1^\diamond, 1^\diamond), q(1^\diamond, 2^\diamond), \dots \rangle. \tag{24}$$

*The interpretation $I$ satisfies every loop formula of every finite loop of $F$ w.r.t. $I$, but it is not a stable model.*

In the example, what distinguishes the set

$$\{p(0^\diamond, 0^\diamond), q(0^\diamond, 1^\diamond), p(1^\diamond, 1^\diamond), q(1^\diamond, 2^\diamond), \dots\} \tag{25}$$

from a loop is that, for every loop contained in (25), there is an outgoing edge in the dependency graph. This is an instance of what we call "unbounded set." Given a dependency graph of $F$ w.r.t. $I$, we say that a nonempty set $Y$ of vertices is *unbounded* w.r.t. $I$ if, for every subset $Z$ of $Y$ that is a loop, there is an edge from a vertex in $Z$ to a vertex in $Y \setminus Z$.

The following proposition tells us how an unbounded set can be characterized by a second-order formula.

**Proposition 4** *Let* $\mathbf{q}$ *be a list of predicate names corresponding to* $\mathbf{p}$*, and let* $Y$ *be a set of atoms in the dependency graph of* $F$ *w.r.t.* $I$ *such that*

$$p_i(\boldsymbol{\xi}^{\diamond}) \in Y \text{ iff } I \models q_i(\boldsymbol{\xi}^{\diamond}),$$

*where* $\boldsymbol{\xi}^{\diamond}$ *is a list of object names. Then*

$$I \models \mathit{Nonempty}(\mathbf{q}) \land \forall \mathbf{v}((\mathbf{v} \leq \mathbf{q}) \land \mathit{Loop}_F(\mathbf{v}) \rightarrow E_F(\mathbf{v}, \mathbf{q}))$$

*iff* $Y$ *is an unbounded set of* $F$ *w.r.t.* $I$*.*

In order to check the stability of a model, we need to check the external support of every loop and every unbounded set. An *extended loop* of $F$ w.r.t. $I$ is a loop or an unbounded set of $F$ w.r.t. $I$. We define $\mathit{Ext\text{-}Loop}_F(\mathbf{u})$ as

$$\mathit{Loop}_F(\mathbf{u}) \lor (\mathit{Nonempty}(\mathbf{u}) \land \ \forall \mathbf{v}((\mathbf{v} \leq \mathbf{u}) \land \mathit{Loop}_F(\mathbf{v}) \rightarrow E_F(\mathbf{v}, \mathbf{u}))). \tag{26}$$

From Propositions 3 and 4, it follows that $I \models \mathit{Ext\text{-}Loop}_F(\mathbf{q})$ iff $Y$ is an extended loop of $F$ w.r.t. $I$.

If we replace $\mathit{Loop}_F(\mathbf{u})$ with $\mathit{Ext\text{-}Loop}_F(\mathbf{u})$ in (23), the formula is equivalent to $\mathrm{SM}[F]$, as the following theorem states.

**Theorem 3** *For any rectified sentence* $F$*, the following sentences are equivalent to each other:*

(*a*) $\mathrm{SM}[F]$;

(*b*) $F \land \forall \mathbf{u}((\mathbf{u} \leq \mathbf{p}) \land \mathit{Nonempty}(\mathbf{u}) \rightarrow \neg \mathit{NSES}_F(\mathbf{u}))$;

(*c*) $F \land \forall \mathbf{u}((\mathbf{u} \leq \mathbf{p}) \land \mathit{Ext\text{-}Loop}_F(\mathbf{u}) \rightarrow \neg \mathit{NSES}_F(\mathbf{u}))$.

In the following example we use the following fact to simplify the formulas.

**Proposition 5** *For any negative formula* $F$*, formula*

$$\mathit{NSES}_F(\mathbf{u}) \leftrightarrow F$$

*is logically valid.*

**Example 2 (continued)** Consider program (7) from Example 2:

$$\begin{array}{l} p(x) \leftarrow q(x) \\ q(y) \leftarrow p(y) \\ p(z) \leftarrow \mathit{not}\ r(z). \end{array}$$

Let $F$ be the FOL-representation of the program:

$$\forall x\big(q(x) \rightarrow p(x)\big) \land \forall y\big(p(y) \rightarrow q(y)\big) \land \forall z\big(\neg r(z) \rightarrow p(z)\big).$$

**1.** SM[$F$] is equivalent to

$$F \wedge \neg\exists u_1u_2u_3((u_1,u_2,u_3) < (p,q,r)) \wedge \\ \forall x(u_2(x) \rightarrow u_1(x)) \wedge \forall y(u_1(y) \rightarrow u_2(y)) \wedge \forall z(\neg r(z) \rightarrow u_1(z))).$$

**2. Formula in Theorem 3 (b):**

$$F \wedge \forall \mathbf{u}(\mathbf{u} \leq \mathbf{p} \wedge \mathit{Nonempty}(\mathbf{u}) \rightarrow \neg \mathit{NSES}_F(\mathbf{u}))$$

is equivalent to

$$\begin{array}{l} F \wedge \forall u_1u_2u_3((u_1,u_2,u_3) \leq (p,q,r) \wedge (\exists x\ u_1(x) \vee \exists x\ u_2(x) \vee \exists x\ u_3(x)) \\ \quad \rightarrow \neg(\forall x[q(x) \wedge \neg u_2(x) \rightarrow p(x) \wedge \neg u_1(x)] \\ \qquad \wedge \forall y[p(y) \wedge \neg u_1(y) \rightarrow q(y) \wedge \neg u_2(y)] \\ \qquad \wedge \forall z[\neg r(z) \rightarrow p(z) \wedge \neg u_1(z)])). \end{array} \tag{27}$$

**3. Formula in Theorem 3 (c):** Similar to (27) except that

$$\exists x\, u_1(x) \vee \exists x\, u_2(x) \vee \exists x\, u_3(x)$$

in (27) is replaced with $\mathit{Ext\text{-}Loop}_F(\mathbf{u})$, which is

$$\begin{array}{l} \mathit{Loop}_F(\mathbf{u}) \vee [(\exists x\, u_1(x) \vee \exists x\, u_2(x) \vee \exists x\, u_3(x)) \\ \wedge\ \forall v_1v_2v_3(((v_1,v_2,v_3) \leq (u_1,u_2,u_3)) \wedge \mathit{Loop}_F(\mathbf{v}) \\ \qquad \rightarrow (\exists x(v_1(x) \wedge u_2(x) \wedge \neg v_2(x)) \vee \exists y(v_2(y) \wedge u_1(y) \wedge \neg v_1(y))))], \end{array}$$

where $\mathit{Loop}_F(\mathbf{u})$ is

$$\begin{array}{l} (\exists x\, u_1(x) \vee \exists x\, u_2(x) \vee \exists x\, u_3(x)) \\ \wedge\ \forall v_1v_2v_3(((\exists x\, v_1(x) \vee \exists x\, v_2(x) \vee \exists x\ v_3(x)) \wedge (v_1,v_2,v_3) < (u_1,u_2,u_3)) \\ \quad \rightarrow (\exists x(v_1(x) \wedge u_2(x) \wedge \neg v_2(x)) \vee \exists y(v_2(y) \wedge u_1(y) \wedge \neg v_1(y)))). \end{array}$$

The proof of Theorem 2 follows from Theorem 3 using the following lemma.

**Lemma 3** *Let $F$ be a rectified sentence of a signature $\sigma$ (possibly containing function constants of positive arity), and let $I$ be an interpretation of $\sigma$ that satisfies $F$. If $F$ is bounded w.r.t. $I$,*

$$I \models \exists \mathbf{u}(\mathbf{u} \leq \mathbf{p} \wedge \mathit{Ext\text{-}Loop}_F(\mathbf{u}) \wedge \mathit{NSES}_F(\mathbf{u}))$$

*iff there is a finite loop $Y$ of $F$ w.r.t. $I$ such that*

$$I \models \Big(\bigwedge Y \wedge \mathit{NES}_F(Y)\Big).$$

## 5. Representing First-Order Stable Model Semantics by First-Order Loop Formulas

We noted in the previous section that if a sentence is bounded w.r.t. a model, then loop formulas can be used to check the stability of the model. In this section, we provide a few syntactic counterparts of the boundedness condition.

### 5.1 Bounded Formulas

We say that a rectified formula $F$ is *bounded* if every infinite path in the first-order dependency graph of $F$ visits only finitely many vertices. If $F$ is bounded, then, clearly, every loop of $F$ is finite. Again, the definition is extended to a non-rectified formula by first rewriting it as a rectified formula. It also applies to a program by referring to its FOL-representation.

One might wonder if the syntactic notion of boundedness ensures the semantic notion of boundedness: that is, if a formula is bounded, then it is bounded w.r.t. any interpretation. However, the following example tells us that this is not the case in general.

**Example 7** *Consider the FOL-representation $F$ of the following program*

$$\begin{array}{l} p(a) \leftarrow q(x) \\ q(x) \leftarrow p(b), \end{array} \tag{28}$$

*and an interpretation $I$ whose universe $|I|$ is the set of all nonnegative integers, $a^I = b^I = 0$, $p^I = \{0\}$ and $q^I = |I|$. Formula (28) is bounded according to the above definition, but not bounded w.r.t. $I$: the dependency graph of $F$ w.r.t. $I$ contains an infinite path such as*

$$\langle p(0^\diamond), q(1^\diamond), p(0^\diamond), q(2^\diamond), \dots \rangle.$$

#### 5.1.1 Bounded Formulas and Clark's Equational Theory

On the other hand, such a relationship holds if the interpretation satisfies Clark's equational theory (1978). Clark's equational theory of a signature $\sigma$, denoted by $\mathrm{CET}_\sigma$, is the union of the universal closures of the following formulas

$$f(x_1, \dots, x_m) \neq g(y_1, \dots, y_n), \tag{29}$$

for all pairs of distinct function constants $f$, $g$,

$$f(x_1, \dots, x_n) = f(y_1, \dots, y_n) \rightarrow (x_1 = y_1 \wedge \dots \wedge x_n = y_n), \tag{30}$$

for all function constants $f$ of arity $> 0$, and

$$t \neq x, \tag{31}$$

where $t$ is any term which contains the variable $x$.

**Proposition 6** *If a rectified formula $F$ of a signature $\sigma$ is bounded, then $F$ is bounded w.r.t. any interpretation of $\sigma$ that satisfies* $\mathrm{CET}_\sigma$.

The following lemma relates loops and loop formulas of different notions of dependency graphs.

**Proposition 7** *For any rectified sentence $F$ of a signature $\sigma$ and for any interpretation $I$ of $\sigma$ that satisfies $\mathrm{CET}_\sigma$, $I$ is a model of*

$$\{LF_F(Y) \mid Y \textit{ is a finite first-order loop of } F\}$$

*iff $I$ is a model of*

$$\{LF_F(Y) \mid Y \textit{ is a finite loop of } F \textit{ w.r.t. } I\}.$$

The following theorem follows from Theorem 2, Proposition 6 and Proposition 7.

**Theorem 4** *Let $F$ be a rectified sentence of a signature $\sigma$ (possibly containing function constants of positive arity), and let $I$ be an interpretation of $\sigma$ that satisfies $F$ and $\mathrm{CET}_\sigma$. If $F$ is bounded, then the following conditions are equivalent to each other:*

*(a)* $I \models \mathrm{SM}[F]$;

*(b) for every nonempty finite set $Y$ of atoms of $\sigma(F)$, $I$ satisfies $LF_F(Y)$;*

*(c) for every finite first-order loop $Y$ of $F$, $I$ satisfies $LF_F(Y)$.*

**Proof**. By Proposition 6, if $F$ is bounded then $F$ is bounded w.r.t. any interpretation that satisfies $\mathrm{CET}_\sigma$. Then the equivalence between (a) and (b) follows from the equivalence between (a) and (b) of Theorem 2. The equivalence between (a) and (c) follows from the equivalence between (a) and (c) of Theorem 2 and by Proposition 7. □

As every Herbrand interpretation of $\sigma$ satisfies $\mathrm{CET}_\sigma$, Theorem 4 applies to Herbrand interpretations as a special case.

The theorem also applies to logic programs, since they can be viewed as a special case of formulas. For example, consider the following program, which is bounded.

$$\begin{array}{l} p(f(x)) \leftarrow q(x) \\ q(x) \leftarrow p(x), r(x) \\ p(a) \\ r(a) \\ r(f(a)). \end{array} \tag{32}$$

The set $\{p(a), p(f(a)), p(f(f(a))), q(a), q(f(a)), r(a), r(f(a))\}$ is an answer set of (32). In accordance with Theorem 4, it is also the Herbrand interpretation of the signature obtained from the program that satisfies the FOL-representation of (32) and the loop formulas, which are the universal closures of

$$\begin{array}{l} p(z) \rightarrow (q(x) \wedge z = f(x)) \vee z = a \\ q(z) \rightarrow p(z) \wedge r(z) \\ r(z) \rightarrow z = a \vee z = f(a). \end{array}$$

Consider another example program by Bonatti (2004), where $\mathtt{a}, \dots, \mathtt{z}$, $\mathtt{nil}$ are object constants.

$$\begin{array}{l} \mathit{letter}(\mathtt{a}) \\ \dots \\ \mathit{letter}(\mathtt{z}) \\ \mathit{atomic}([x]) \leftarrow \mathit{letter}(x) \\ \mathit{atomic}([x|y]) \leftarrow \mathit{letter}(x), \mathit{atomic}(y). \end{array} \tag{33}$$

The expression $[x|y]$ is a list whose head is $x$ and whose tail is $y$, which stands for a function $\mathit{cons}(x, y)$. The expression $[x]$ stands for $\mathit{cons}(x, \mathtt{nil})$ where $\mathtt{nil}$ is a special symbol for the empty list. This program is bounded. The only answer set of the program is the only Herbrand interpretation of the FOL-representation of (33) and the universal closures of

$$\begin{aligned} &letter(u) \rightarrow u = \mathtt{a} \vee \ldots \vee u = \mathtt{z} \\ &atomic(u) \rightarrow \exists v\ (letter(v) \wedge u = cons(v, nil)) \\ &\qquad\qquad \vee\ \exists xy\ (letter(x) \wedge atomic(y) \wedge y \neq u \wedge u = cons(x, y)). \end{aligned}$$

In fact, the definitions of standard list processing predicates, such as `member`, `append`, and `reverse` (Bonatti, 2004, Figure 1) are bounded, so they can be represented by first-order formulas on Herbrand interpretations.[12]

We say that a formula $F$ is *atomic-tight* if the first-order dependency graph of $F$ has no infinite paths. Every tight sentence is atomic-tight, but not vice versa. For example, the FOL-representations of programs (32) and (33) are atomic-tight, but are not tight. Similar to Proposition 6, if $F$ is atomic-tight, then $F$ is atomic-tight w.r.t. any interpretation that satisfies $\text{CET}_\sigma$, so that the following statement is derived from Corollary 3.

**Corollary 5** *Let $F$ be a rectified sentence of a signature $\sigma$ (possibly containing function constants of positive arity), and let $I$ be an interpretation of $\sigma$ that satisfies $F$ and* $\text{CET}_\sigma$. *If $F$ is atomic-tight, then $I$ satisfies* $\text{SM}[F]$ *iff $I$ satisfies* $\text{SLF}[F]$.

The statement of Corollary 5 is restricted to interpretations that satisfy $\text{CET}_\sigma$. Indeed, the statement becomes wrong if this restriction is dropped. For example, program (28) in Example 7 is atomic-tight, but the non-stable model considered there satisfies all loop formulas, including those of singleton loops.

#### 5.1.2 Bounded Formulas and Normal Form

Normal form is another syntactic condition that can be imposed so that the syntactic notion of boundedness ensures the semantic notion of boundedness. We say that a formula is in *normal form* if every strictly positive occurrence of an atom is of the form $p(\mathbf{x})$, where $\mathbf{x}$ is a list of distinct variables. It is clear that every formula can be turned into normal form using equality.

**Proposition 8** *If a rectified formula $F$ in normal form is bounded, then $F$ is bounded w.r.t. any interpretation.*

**Proposition 9** *If a rectified sentence $F$ in normal form is bounded, then for any interpretation $I$, $I$ is a model of*

$$\{LF_F(Y) \mid Y \text{ is a finite first-order loop of } F\}$$

*iff $I$ is a model of*

$$\{LF_F(Y) \mid Y \text{ is a finite loop of } F \text{ w.r.t. } I\}.$$

The following theorem follows from Theorem 2, Proposition 8 and Proposition 9.

12. They actually satisfy a stronger condition called "finitely recursive" (Bonatti, 2004). See Section 8 for more details.

**Theorem 5** *Let $F$ be a rectified sentence in normal form (possibly containing function constants of positive arity). If $F$ is bounded, then the following formulas are equivalent to each other:*

*(a)* $\mathrm{SM}[F]$*;*

*(b)* $\{F\} \cup \{LF_F(Y) \mid Y$ *is a nonempty finite set of atoms of* $\sigma(F)\}$*;*

*(c)* $\{F\} \cup \{LF_F(Y) \mid Y$ *is a finite first-order loop of* $F\}$.

**Proof**. By Proposition 8, if $F$ is bounded then $F$ is bounded w.r.t. any interpretation $I$. Then the equivalence between (a) and (b) follows from the equivalence between (a) and (b) of Theorem 2. The equivalence between (a) and (c) follows from the equivalence between (a) and (c) of Theorem 2 and by Proposition 9. □

Consider a program in normal form

$$\begin{array}{l} p(x) \leftarrow x = a,\ q(a) \\ q(y) \leftarrow p(b) \end{array} \tag{34}$$

and an interpretation $I$ such that $|I| = \{1\}$, $a^I = b^I = 1$ and $p^I = q^I = \{1\}$. This interpretation does not satisfy Clark's equational theory, and is not a stable model. In accordance with Theorem 5, $I$ does not satisfy the loop formula of the loop $\{p(b), q(a)\}$, which is

$$p(b) \wedge q(a) \rightarrow (b = a \wedge q(a) \wedge a \neq a) \vee (p(b) \wedge b \neq b).$$

On the other hand, consider another program in *non-normal* form that has the same stable models as (34):

$$\begin{array}{l} p(a) \leftarrow q(a) \\ q(y) \leftarrow p(b) \end{array} \tag{35}$$

Program (35) has a finite complete set of loops, $\{\{p(z)\}, \{q(z)\}\}$; their loop formulas are the universal closures of

$$\begin{array}{l} p(z) \rightarrow z = a \wedge q(a) \\ q(z) \rightarrow p(b) \end{array}$$

and $I$ satisfies all loop formulas. This example illustrates the role of normal form assumption in Theorem 5 (in place of Clark's equational theory in Theorem 4).

Note that a normal form conversion may turn a bounded sentence into a non-bounded sentence. For instance, the normal form of the bounded program (32) is

$$\begin{array}{l} p(y) \leftarrow y = f(x), q(x) \\ q(x) \leftarrow p(x), r(x) \\ p(x) \leftarrow x = a \\ r(x) \leftarrow x = a \\ r(x) \leftarrow x = f(a), \end{array} \tag{36}$$

which is not bounded.

Unlike in Corollary 5, if a program is in normal form, atomic-tightness is not more general than tightness. It is not difficult to check that a program in normal form is atomic-tight iff it is tight.

#### 5.1.3 Decidability of Boundedness and Finite Complete Set of Loops

In general, checking whether $F$ is bounded is not decidable, but it becomes decidable if $F$ contains no function constants of positive arity. The same is the case for checking whether $F$ is atomic-tight.

**Proposition 10** *For any rectified sentence $F$ (allowing function constants of positive arity),*

*(a) checking whether $F$ is bounded is not decidable;*

*(b) checking whether $F$ is atomic-tight is not decidable.*

*If $F$ contains no function constants of positive arity,*

*(c) checking whether $F$ is bounded is decidable;*

*(d) checking whether $F$ is atomic-tight is decidable.*

The proof of Proposition 10 (c) is based on the following fact and the straightforward extension of Theorem 2 by Chen et al. (2006) to first-order formulas, which asserts that checking if $F$ has a finite complete set of loops is decidable.

**Proposition 11** *For any rectified formula $F$ that contains no function constants of positive arity, $F$ is bounded iff $F$ has a finite complete set of loops.*

Note that Proposition 11 does not hold if $F$ is allowed to contain function constants of positive arity. For instance,

$$p(x) \leftarrow p(f(x))$$

is not bounded, but has a finite complete set of loops $\{\{p(x)\}\}$.

The following corollary follows from Theorem 4 and Proposition 11.

**Corollary 6** *Let $F$ be a rectified sentence of a signature $\sigma$ that has no function constants of positive arity, and let $I$ be an interpretation of $\sigma$ that satisfies $F$ and $\mathrm{CET}_\sigma$. If $F$ has a finite complete set of loops, then conditions (a), (b), and (c) of Theorem 4 are equivalent to each other.*

The following corollary follows from Theorem 5 and Proposition 11.

**Corollary 7** *Let $F$ be a rectified sentence in normal form that has no function constants of positive arity. If $F$ has a finite complete set of loops, formulas in (a), (b), and (c) of Theorem 5 are equivalent to each other.*

### 5.2 Semi-Safe Formulas

Semi-safety is another decidable syntactic condition that ensures that SM$[F]$ can be expressed by first-order sentences.

We assume that there are no function constants of positive arity. According to Lee, Lifschitz, and Palla (2009), a semi-safe sentence has the "small predicate property": the relation represented by any of its predicate constants $p$ can hold for a tuple of arguments only if each member of the tuple is represented by an object constant occurring in $F$. We will show that any semi-safe sentence under the stable model semantics can be turned into a sentence in first-order logic.

First, we review the notion of semi-safety by Lee et al. (2009).[13] As a preliminary step, we assign to every formula $F$ a set RV$(F)$ of its *restricted variables* as follows:

- For an atomic formula $F$,
  - if $F$ is an equality between two variables, then $\text{RV}(F) = \emptyset$;
  - otherwise, RV$(F)$ is the set of all variables occurring in $F$;
- $\text{RV}(G \wedge H) = \text{RV}(G) \cup \text{RV}(H)$;
- $\text{RV}(G \vee H) = \text{RV}(G) \cap \text{RV}(H)$;
- $\text{RV}(G \rightarrow H) = \emptyset$;
- $\text{RV}(QvG) = \text{RV}(G) \setminus \{v\}$ where $Q \in \{\forall, \exists\}$.

We say that a variable $x$ is *restricted* in $F$ if $x \in \text{RV}(F)$. A rectified formula $F$ is *semi-safe* if every strictly positive occurrence of every variable $x$ belongs to a subformula $G \rightarrow H$ where $x$ is restricted in $G$.

If a sentence has no strictly positive occurrence of a variable, then it is obviously semi-safe. The FOL-representation of a disjunctive program is semi-safe if, for each rule (10) of the program, every variable occurring in the head of the rule occurs in $B$ as well.

**Example 8** *The FOL-representation of (8) is not semi-safe. Formula*

$$p(a) \wedge q(b) \wedge \forall xy((p(x) \vee q(y)) \rightarrow p(y))$$

*is not semi-safe, while*

$$p(a) \wedge q(b) \wedge \forall xy((p(x) \wedge q(y)) \rightarrow p(y)) \tag{37}$$

*is semi-safe.*

For any finite set $\mathbf{c}$ of object constants, $in_{\mathbf{c}}(x)$ stands for the formula

$$\bigvee_{c \in \mathbf{c}} x = c.$$

13. The definition here is slightly more general in that it does not refer to prenex form. Instead we require a formula to be rectified.

The small predicate property can be expressed by the conjunction of the sentences

$$\forall v_1, \ldots, v_n \Big( p(v_1, \ldots, v_n) \rightarrow \bigwedge_{i=1,\ldots,n} in_{\mathbf{c}}(v_i) \Big)$$

for all predicate constants $p$ occurring in $F$, where $v_1, \ldots, v_n$ are distinct variables. We denote this conjunction of the sentences by $SPP_{\mathbf{c}}$. By $\mathbf{c}(F)$ we denote the set of all object constants occurring in $F$.

**Proposition 12** *(Lee et al., 2009) For any semi-safe sentence $F$, formula* SM$[F]$ *entails $SPP_{\mathbf{c}(F)}$.*

For example, for the semi-safe sentence (37), SM[(37)] entails

$$\forall x \Big( p(x) \rightarrow (x = a \vee x = b)) \wedge \forall x (q(x) \rightarrow (x = a \vee x = b) \Big). \tag{38}$$

The following proposition tells us that for a semi-safe sentence $F$, formula SM$[F]$ can be equivalently rewritten as a first-order sentence.

**Theorem 6** *Let $F$ be a rectified sentence that has no function constants of positive arity. If $F$ is semi-safe, then* SM$[F]$ *is equivalent to the conjunction of $F$, $SPP_{\mathbf{c}(F)}$ and a finite number of first-order loop formulas.*

**Proof**. If $F$ is semi-safe, then SM$[F]$ entails $SPP_{\mathbf{c}(F)}$. So it is sufficient to prove that under the assumption $SPP_{\mathbf{c}(F)}$, SM$[F]$ is equivalent to the conjunction of $F$ and a finite number of first-order loop formulas. It follows from $I \models SPP_{\mathbf{c}(F)}$ that $F$ is bounded w.r.t. $I$. Since every finite loop of $F$ w.r.t. $I$ can be represented by a finite set of atoms whose terms are object variables, it follows from Theorem 2 that $I$ satisfies SM$[F]$ iff $I$ satisfies the loop formulas of those sets. □

For example, SM[(37)] is equivalent to the conjunction of $F$, (38) and the universal closures of

$$\begin{array}{lcl} p(z) & \rightarrow & z = a \ \vee \ (p(x) \wedge q(z) \wedge z \neq x) \\ q(z) & \rightarrow & z = b \end{array}$$

Note that the condition on a finite complete set of loops in Corollaries 6 and 7, and the condition on semi-safety in Theorem 6 do not entail each other. For instance, formula (37) is semi-safe, but has no finite complete set of first-order loops, while $\exists x\, p(x)$ has a finite complete set of loops $\{\{p(x)\}\}$, but it is not semi-safe. Also program $\Pi_1$ in Section 1 has a finite complete set of loops, but it is not semi-safe due to $w$ in the fourth rule.

## 6. Programs with Explicit Quantifiers

In the following we extend the syntax of a logic program by allowing explicit quantifiers. A *rule with quantifiers* is of the form

$$H \leftarrow G, \tag{39}$$

where $G$ and $H$ are first-order formulas such that every occurrence of every implication in $G$ and $H$ belongs to a negative formula. A *program with quantifiers* is a finite set of rules with quantifiers. Program $\Pi_1$ in Section 1 is an example. The semantics of such a program is defined by identifying the program with its FOL-representation under the stable model semantics. By restricting the syntax of a program like the one above, in comparison with the syntax of an arbitrary formula, we are able to write a more succinct loop formulas, as we show below.

Let $F$ be a formula and $Y$ a finite set of atoms. By $F_Y$ we denote the formula obtained from $F$ by replacing every occurrence of every atom $p(\mathbf{t})$ in $F$ that does not belong to a negative formula with $p(\mathbf{t}) \wedge \bigwedge_{p(\mathbf{t}')\in Y} \mathbf{t} \neq \mathbf{t}'$. Let $\Pi$ be a program with quantifiers. Given a finite set $Y$ of atoms of $\sigma(\Pi)$, we first rename variables in $\Pi$ so that no variables in $\Pi$ occur in $Y$. We define the formula $QES_\Pi(Y)$ ("*External Support Formula for Programs with Quantifiers*") to be the disjunction of

$$\exists \mathbf{z}(G_Y \wedge \neg H_Y) \tag{40}$$

for every rule (39) such that $H$ contains a strictly positive occurrence of a predicate constant that occurs in $Y$, and $\mathbf{z}$ is the list of all free variables in the rule that do not occur in $Y$.

The loop formula of $Y$ for $\Pi$ is the universal closure of

$$\bigwedge Y \rightarrow QES_\Pi(Y). \tag{41}$$

The following proposition tells us that (41) is equivalent to (12) when the notions are applied to a program with explicit quantifiers. It also shows that (41) is a generalization of the definition of a loop formula for a disjunctive program.

**Proposition 13** *Let $\Pi$ be a program with quantifiers, $F$ the FOL-representation of $\Pi$, and $Y$ a finite set of atoms. Under the assumption $\Pi$, formula $QES_\Pi(Y)$ is equivalent to $\neg NES_F(Y)$. If $\Pi$ is a disjunctive program in normal form, then $QES_\Pi(Y)$ is also equivalent to $ES_\Pi(Y)$ under the assumption $\Pi$.*

Note that the size of (41) for each $Y$ is polynomial to the size of the given program. This is not the case when we apply (12) to the FOL-representation of the program, due to the expansion of *NES* for nested implications. On the other hand, the syntactic condition imposed on the rule with quantifiers avoids such an exponential blow up, as the following lemma tells us.

**Lemma 4** *Let $F$ be a formula such that every occurrence of an implication in $F$ belongs to a negative formula and let $Y$ be a set of atoms. $NES_F(Y)$ is equivalent to $F_Y$.*

**Proof**. By induction on $F$. □

**Example 2 (continued) First-Order Loop Formula when $\Pi$ is understood as an extended program (Using $QES_\Pi(Y)$) :** *Under the assumption $\Pi$,*

- *$LF_\Pi(Y_1)$ is equivalent to the universal closure of*

$$p(u) \rightarrow (\exists x(q(x) \wedge \neg(p(x) \wedge x \neq u)) \vee \ \exists z(\neg r(z) \wedge \neg(p(z) \wedge z \neq u))).$$

- $LF_\Pi(Y_2)$ *is equivalent to the universal closure of*

$$q(u) \rightarrow \exists y(p(y) \wedge \neg(q(y) \wedge y \neq u)).$$

- $LF_\Pi(Y_3)$ *is equivalent to the universal closure of*

$$r(u) \rightarrow \bot.$$

- $LF_\Pi(Y_4)$ *is equivalent to the universal closure of*

$$\begin{aligned}(p(u) \wedge q(u)) \rightarrow (&\exists x((q(x) \wedge x \neq u) \wedge \neg(p(x) \wedge x \neq u)) \\ &\vee \exists y((p(y) \wedge y \neq u) \wedge \neg(q(y) \wedge y \neq u)) \\ &\vee \exists z(\neg r(z) \wedge \neg(p(z) \wedge z \neq u))).\end{aligned}$$

A finite set $\Gamma$ of sentences *entails* a sentence $F$ under the stable model semantics (symbolically, $\Gamma \models_{\rm SM} F$), if every stable model of $\Gamma$ satisfies $F$.

If SM$[F]$ can be reduced to a first-order sentence, as described in Theorem 5 and Theorem 6, then

$$\Gamma \models_{\rm SM} F \text{ iff } \Gamma \cup \Delta \models F,$$

where $\Delta$ is the set of first-order loop formulas required (and possibly including $SPP_{\mathbf{c}(F)}$ when Theorem 6 is applied). This fact allows us to use first-order theorem provers to reason about query entailment under the stable model semantics.

**Example 9** *Consider program* $\Pi_1$ *in Section 1, which has the following finite complete set of loops:* $\{Man(u)\}$, $\{Spouse(u,v)\}$, $\{HasWife(u)\}$, $\{Married(u)\}$, $\{Accident(u,v)\}$, $\{Discount(u,v)\}$, *and* $\{HasWife(u), Married(u)\}$. *Their loop formulas for* $\Pi_1 \cup \Pi_2 \cup \Pi_3$ *are equivalent to the universal closure of*

$$Man(u) \rightarrow \neg\big(Man(John) \wedge John \neq u\big);$$

$$Spouse(u,v) \rightarrow \neg\big(\exists y\big(Spouse(John,y) \wedge (John,y) \neq (u,v)\big)\big);$$

$$\begin{aligned}HasWife(u) \rightarrow\ &\exists x\big(\exists y\, Spouse(x,y) \wedge \neg(HasWife(x) \wedge x \neq u)\big) \\ &\vee \exists x\big(Man(x) \wedge Married(x) \wedge \neg(HasWife(x) \wedge x \neq u)\big);\end{aligned}$$

$$Married(u) \rightarrow \exists x\big(Man(x) \wedge HasWife(x) \wedge \neg(Married(x) \wedge x \neq u)\big);$$

$$Accident(u,v) \rightarrow \bot;$$

$$\begin{aligned}&Discount(u,v) \rightarrow \\ &\quad \exists x\big(Married(x) \wedge \neg\exists z\, Accident(x,z) \wedge \neg(\exists w(Discount(x,w) \wedge (x,w) \neq (u,v)))\big);\end{aligned}$$

$$\begin{aligned}&Married(u) \wedge HasWife(u) \rightarrow \\ &\qquad \exists x\big(\exists y\, Spouse(x,y) \wedge \neg(HasWife(x) \wedge (x \neq u))\big) \\ &\qquad \vee \exists x\big(Man(x) \wedge Married(x) \wedge x \neq u \wedge \neg(HasWife(x) \wedge x \neq u)\big) \\ &\qquad \vee \exists x\big(Man(x) \wedge HasWife(x) \wedge x \neq u \wedge \neg(Married(x) \wedge x \neq u)\big).\end{aligned}$$

*These loop formulas, conjoined with the FOL-representation of* $\Pi_1 \cup \Pi_2 \cup \Pi_3$*, entail under first-order logic each of* $\exists x\ Married(x)$ *and* $\forall xy(Discount(x,y) \rightarrow x = John)$*. We verified the answers using a first-order theorem prover Vampire* [14]*.*

## 7. Extension to Allow Extensional Predicates

The definition of a stable model in the journal paper by Ferraris et al. (2011), reviewed in Section 2, is more general than the definition in their conference paper (Ferraris et al., 2007) in that it allows us to distinguish between intensional and non-intensional (a.k.a. extensional) predicates. Similar to Datalog, intensional (output) predicates are characterized in terms of extensional (input) predicates. For instance, consider Example 9 again, but now assume that *Man* and *Spouse* are non-intensional. $\Pi_1 \cup \Pi_2 \cup \Pi_3$ still entails $\exists xyDiscount(x,y)$ but no longer entails $\forall xy(Discount(x,y) \rightarrow x = John)$ because there may be a person other than *John* who has a spouse.

The results in the earlier sections can be extended to this general semantics in view of Proposition 14 below, which characterizes $\mathrm{SM}[F;\mathbf{p}]$ in terms of $\mathrm{SM}[F]$. By $pr(F)$ we denote the list of all predicate constants occurring in $F$; by $Choice(\mathbf{p})$ we denote the conjunction of "choice formulas" $\forall\mathbf{x}(p(\mathbf{x}) \vee \neg p(\mathbf{x}))$ for all predicate constants $p$ in $\mathbf{p}$, where $\mathbf{x}$ is a list of distinct object variables; by $False(\mathbf{p})$ we denote the conjunction of $\forall\mathbf{x}\neg p(\mathbf{x})$ for all predicate constants $p$ in $\mathbf{p}$. We sometimes identify a list with the corresponding set when there is no confusion.

**Proposition 14** *For any list* $\mathbf{p}$ *of predicate constants, formula* $\mathrm{SM}[F;\mathbf{p}]$ *is equivalent to*

$$\mathrm{SM}[F \wedge Choice(pr(F)\backslash\mathbf{p}) \wedge False(\mathbf{p}\backslash pr(F))] \tag{42}$$

*and to*

$$\mathrm{SM}[F^{\neg\neg} \wedge Choice(pr(F)\backslash\mathbf{p}) \wedge False(\mathbf{p}\backslash pr(F))], \tag{43}$$

*where* $F^{\neg\neg}$ *is obtained from* $F$ *by replacing every atom of the form* $q(\mathbf{t})$ *in* $F$ *such that* $q$ *does not belong to* $\mathbf{p}$ *by* $\neg\neg q(\mathbf{t})$*.*

This proposition allows us to extend the results established for $\mathrm{SM}[F]$ to $\mathrm{SM}[F;\mathbf{p}]$. For instance, Theorem 3 can be extended to $\mathrm{SM}[F;\mathbf{p}]$ by first rewriting it into the form $\mathrm{SM}[G]$, where $G$ is

$$F^{\neg\neg} \wedge Choice(pr(F)\backslash\mathbf{p}) \wedge False(\mathbf{p}\backslash pr(F)). \tag{44}$$

In the next three corollaries, $\sigma$ is a signature, $F$ is a rectified sentence of $\sigma$ (possibly containing function constants of positive arity), $\mathbf{p}$ is any finite list of predicate constants from $\sigma$, and $G$ is (44).

The first corollary follows from Theorem 2 and Proposition 14.

**Corollary 8** *For any interpretation* $I$ *of* $\sigma$ *that satisfies* $F$*, if* $G$ *is bounded w.r.t.* $I$*, then the following conditions are equivalent to each other:*

*(a)* $I \models \mathrm{SM}[F;\mathbf{p}]$*;*

14. `http://www.vampire.fm` .

(b) *for every nonempty finite set $Y$ of atoms formed from predicate constants in* $\mathbf{p}$ *and object names for* $|I|$, *$I$ satisfies* $LF_F(Y)$;

(c) *for every finite loop $Y$ of $G$ w.r.t. $I$ whose predicate constants are contained in* $\mathbf{p}$, *$I$ satisfies* $LF_F(Y)$.

The next corollary follows from Theorem 4 and Proposition 14.

**Corollary 9** *If $G$ is bounded, then, for any interpretation $I$ of $\sigma$ that satisfies $F$ and* $\mathrm{CET}_\sigma$, *the following conditions are equivalent to each other:*

(a) $I \models \mathrm{SM}[F; \mathbf{p}]$;

(b) *for every nonempty finite set $Y$ of atoms of $\sigma(G)$ whose predicate constants are contained in* $\mathbf{p}$, *$I$ satisfies* $LF_F(Y)$;

(c) *for every finite first-order loop $Y$ of $G$ whose predicate constants are contained in* $\mathbf{p}$, *$I$ satisfies* $LF_F(Y)$.

The last corollary follows from Theorem 5 and Proposition 14.

**Corollary 10** *If $G$ is in normal form and is bounded, then the following formulas are equivalent to each other:*

(a) $\mathrm{SM}[F; \mathbf{p}]$;

(b) $\{F\} \cup \{LF_F(Y) \mid Y$ *is a nonempty finite set of atoms of $\sigma(G)$ whose predicate constants are contained in* $\mathbf{p}\}$;

(c) $\{F\} \cup \{LF_F(Y) \mid Y$ *is a finite first-order loop of $G$ whose predicate constants are contained in* $\mathbf{p}\}$.

**Example 10** *Consider Example 9 again, assuming that Man and Spouse are extensional. Let $F$ be the FOL-presentation of $\Pi_1 \cup \Pi_2 \cup \Pi_3$ and let $G$ be the formula (44). The loops of $G$ are the same as the loops of $F$. The loop formulas remain the same as before except for the following loop formulas of $Man(u)$ and $Spouse(u, v)$:*

$$Man(u) \rightarrow \neg\big(Man(John) \wedge John \neq u\big) \vee \exists x\, \neg\big((Man(x) \wedge x \neq u) \vee \neg Man(x)\big);$$

$$\begin{aligned} Spouse(u, v) \rightarrow \neg\big(&\exists y\big(Spouse(John, y) \wedge (John, y) \neq (u, v)\big)\big) \vee \\ &\exists xy\, \neg\big((Spouse(x, y) \wedge (x, y) \neq (u, v)) \vee \neg Spouse(x, y)\big). \end{aligned}$$

*These two formulas are tautologies. As a result, the loop formulas of all loops, conjoined with $G$, entail $\exists xy Discount(x, y)$, but no longer entail $\forall xy\ (Discount(x, y) \rightarrow x = John)$.*

In general, there are no loops of $G$ that contain both intensional and extensional predicates. Also every loop of $G$ that contains an extensional predicate is a singleton, and the loop formula of such a loop is a tautology.

Corollary 3 is extended to allow extensional predicates as in the following. By $\text{SLF}[F;\mathbf{p}]$, we denote

$$\{LF_F(\{p(\mathbf{x})\}) \mid p \text{ is a predicate constant in } \mathbf{p}, \text{ and } \mathbf{x} \text{ is a list of distinct object variables whose length is the same as the arity of } p\}.$$

We say that a formula $F$ is $\mathbf{p}$-*atomic-tight* w.r.t. $I$ if every infinite path in the dependency graph of $F$ w.r.t. $I$ whose vertices are satisfied by $I$ contains an atom whose predicate constant is not in $\mathbf{p}$.

**Corollary 11** *Let $F$ be a rectified sentence (possibly containing function constants of positive arity), and let $I$ be a model of $F$. If $F$ is $\mathbf{p}$-atomic-tight w.r.t. $I$, then $I$ satisfies* $\text{SM}[F;\mathbf{p}]$ *iff $I$ satisfies* $\text{SLF}[F;\mathbf{p}]$.

The definition of semi-safety is extended to distinguish between intensional and non-intensional predicates as follows. Let $F$ be a formula that has no function constants of positive arity. To every first-order formula $F$ we assign a set $\text{RV}_{\mathbf{p}}(F)$ of *restricted variables relative to* $\mathbf{p}$ as follows.

- For an atomic formula $F$ (including equality and $\bot$),
  - if $F$ is an equality between two variables, or is an atom whose predicate constant is not in $\mathbf{p}$, then $\text{RV}_{\mathbf{p}}(F) = \emptyset$;
  - otherwise, $\text{RV}_{\mathbf{p}}(F)$ is the set of all variables occurring in $F$;
- $\text{RV}_{\mathbf{p}}(G \wedge H) = \text{RV}_{\mathbf{p}}(G) \cup \text{RV}_{\mathbf{p}}(H)$;
- $\text{RV}_{\mathbf{p}}(G \vee H) = \text{RV}_{\mathbf{p}}(G) \cap RV_{\mathbf{p}}(H)$;
- $\text{RV}_{\mathbf{p}}(G \rightarrow H) = \emptyset$.
- $\text{RV}_{\mathbf{p}}(QvG) = \text{RV}_{\mathbf{p}}(G) \setminus \{v\}$ where $Q \in \{\forall, \exists\}$.

We say that a variable $x$ is $\mathbf{p}$-*restricted* in $F$ if $x \in \text{RV}_{\mathbf{p}}(F)$. A rectified formula $F$ is *semi-safe relative to* $\mathbf{p}$ if every strictly positive occurrence of every variable $x$ belongs to a subformula $G \rightarrow H$, where $x$ is $\mathbf{p}$-restricted in $G$.

The small predicate property is generalized as follows. Formula $SPP^{\mathbf{p}}_{\mathbf{c}}$ is the conjunction of the sentences

$$\forall v_1, \ldots, v_n \Big(p(v_1, \ldots, v_n) \rightarrow \bigwedge_{i=1,\ldots,n} in_{\mathbf{c}}(v_i)\Big)$$

for all predicate constants $p$ in $\mathbf{p}$, where $v_1, \ldots, v_n$ are distinct variables.

**Proposition 15** *(Lee et al., 2009) For any semi-safe sentence $F$ relative to $\mathbf{p}$, formula* $\text{SM}[F;\mathbf{p}]$ *entails* $SPP^{\mathbf{p}}_{\mathbf{c}(F)}$.

The following proposition tells us that for a semi-safe sentence $F$, formula $\text{SM}[F;\mathbf{p}]$ can be equivalently rewritten as a first-order sentence.

**Theorem 7** *Let $F$ be a rectified sentence that has no function constants of positive arity. If $F$ is semi-safe relative to* $\mathbf{p}$*, then* $\mathrm{SM}[F;\mathbf{p}]$ *is equivalent to the conjunction of $F$,* $SPP^{\mathbf{p}}_{\mathbf{c}(F)}$ *and a finite number of first-order loop formulas.*

**Proof**. Let $F$ be a sentence of the signature $\sigma$. If $F$ is semi-safe relative to $\mathbf{p}$, then $\mathrm{SM}[F;\mathbf{p}]$ entails $SPP^{\mathbf{p}}_{\mathbf{c}(F)}$, so it is sufficient to prove that under the assumption $SPP^{\mathbf{p}}_{\mathbf{c}(F)}$, $\mathrm{SM}[F;\mathbf{p}]$ is equivalent to the conjunction of $F$ and a finite number of first-order loop formulas. By Proposition 14, $\mathrm{SM}[F;\mathbf{p}]$ is equivalent to $\mathrm{SM}[G]$, where $G$ is (44). Consider any interpretation $I$ of $\sigma$ that satisfies $G$ and $SPP^{\mathbf{p}}_{\mathbf{c}(F)}$. Note that the dependency graph of $G$ w.r.t. $I$ contains no outgoing edges from a vertex whose predicate constant does not belong to $\mathbf{p}$. Together with the fact that $I \models SPP^{\mathbf{p}}_{\mathbf{c}(F)}$, we conclude that each path in the dependency graph whose vertices are satisfied by $I$ visits only finitely many vertices. Consequently, $G$ is bounded w.r.t. $I$. Since every finite loop of $G$ w.r.t. $I$ can be represented by a finite set of atoms whose terms are object variables, it follows from Theorem 2 that $I$ satisfies $\mathrm{SM}[G]$ iff $I$ satisfies the loop formulas of those sets. □

## 8. Related Work

The notion of a bounded program is related to the notion of a *finitely recursive* program studied by Bonatti (2004), where a different definition of a dependency graph was considered. The *atom dependency graph* of a nondisjunctive ground program defined by Bonatti is a directed graph such that the vertices are the set of ground atoms, and the edges go from the atom in the head to atoms in the body of every rule, including those in the negative body. A program is called *finitely recursive* if, for every atom, there are only finitely many atoms reachable from it in the atom dependency graph. It is clear that every finitely recursive program is bounded, but the converse does not hold. For instance, the program

$$p(x) \leftarrow not\ p(f(x))$$

is bounded, but is not finitely recursive because there are infinite paths that involve negative edges. Also the program

$$p(a) \leftarrow q(f(x))$$

is bounded, but is not finitely recursive because infinitely many atoms $q(f(a)), q(f(f(a))), \dots$ can be reached from $p(a)$ in the atom dependency graph. Like bounded programs, checking finitely recursive programs is undecidable in the presence of function constants of positive arity.

Lin and Wang (2008) extended answer set semantics with functions by extending the definition of a reduct, and also provided loop formulas for such programs. We can provide an alternative account of their results by considering the notions there as special cases of the definitions presented in this paper. For simplicity, we assume non-sorted languages.[15] Essentially, they restricted attention to a special case of non-Herbrand interpretations such that object constants form the universe, and ground terms other than object constants are mapped to object constants. According to Lin and Wang, an *LW-program* $P$ consists of

15. Lin and Wang (2008) consider essentially many-sorted languages. The result of this section can be extended to that case by considering many-sorted SM (Kim, Lee, & Palla, 2009).

*type definitions* and a set of rules. Type definitions introduce the domains for a many-sorted signature consisting of some object constants, and includes the evaluation of each function symbol of positive arity that maps a list of object constants to an object constant. Since we assume non-sorted languages, we consider only a single domain (universe). We say that an interpretation $I$ is a $P$*-interpretation* if the universe is the set of object constants specified by $P$, object constants are evaluated to itself, and ground terms other than object constants are evaluated conforming to the type definitions of $P$.

**Proposition 16** *Let $P$ be an LW-program and let $F$ be the FOL-representation of the set of rules in $P$. The following conditions are equivalent to each other:*

*(a) $I$ is an answer set of $P$ according to Lin and Wang (2008);*

*(b) $I$ is a $P$-interpretation that satisfies* $\mathrm{SM}[F]$*;*

*(c) $I$ is a $P$-interpretation that satisfies $F$ and the loop formulas of $Y$ for all loops $Y$ of $F$ w.r.t. $I$.*

The equivalence between (b) and (c) follows from Proposition 2 since the universe is finite. The equivalence between (a) and (c) follows from the fact that LW answer sets can be characterized by loop formulas that are defined by Lin and Wang (2008) and that these loop formulas are essentially the same as the loop formulas in (c).

Since the proposal of the first-order stable model semantics, there have been some papers about first-order definability of $\mathrm{SM}[F]$. Zhang and Zhou (2010) show that, for a nondisjunctive program $\Pi$ that has no function constants of positive arity, its first-order stable model semantics can be reformulated by a progression based semantics. They also showed that the programs whose answer sets can be found by a finite progression are exactly those that can be represented by first-order formulas. Some researchers have paid special attention to first-order definability of $\mathrm{SM}[F]$ on finite structures. Chen, Zhang, and Zhou (2010) show a game-theoretic characterization for the first-order indefinability of first-order answer set programs on finite structures. Asuncion, Lin, Zhang, and Zhou (2010) show first-order definability on finite structures by turning programs into modified completion using new predicates to record levels. Chen, Lin, Zhang, and Zhou (2011) present a condition called "loop-separable," which is more refined than finite complete set of loops under which the finite answer sets of a program can be captured by first-order sentences. However, like the condition of finite complete set of loops, this condition is disjoint with semi-safety. The following program is semi-safe but not loop-separable:

$$p(x) \leftarrow p(y), q(x, y).$$

However, all this work is limited to nondisjunctive programs that contain no function constants of positive arity. Our work is not limited to finite structures, and considers function constants of positive arity as well. Nonetheless the above papers on first-order definability are closely related to our work and more insights would be gained from the relationship between them.

The use of first-order theorem provers for the stable model semantics was already investigated by Sabuncu and Alpaslan (2007), but their results are limited in several ways. They considered nondisjunctive logic programs with "trivial" loops only, in which case the stable model semantics is equivalent to the completion semantics. They also restricted attention to Herbrand models.

## 9. Conclusion

This paper puts first-order loop formulas in the context of first-order reasoning and studies how they are related to first-order stable model semantics. The similarities and mismatches found in this paper provide useful insights into first-order reasoning with stable models. Future work is to find further restrictions that make first-order stable model reasoning decidable and computable in an efficient manner, like the conditions imposed in "finitary" programs (Bonatti, 2004). Recently, the first-order stable model semantics was shown to be used as a unifying nonmonotonic logic for integrating rules and ontologies (de Bruijn, Pearce, Polleres, & Valverde, 2010; Lee & Palla, 2011), in which ontology predicates are identified with extensional predicates. Based on the studied relationship between first-order stable model semantics and first-order loop formulas, one may find further restrictions that are tailored to the hybrid knowledge bases for efficient computation.

## Acknowledgments

We are grateful to Joseph Babb, Michael Bartholomew, Piero Bonatti, Vladimir Lifschitz, and Ravi Palla for useful discussions, and to the anonymous referees for their useful comments. The authors were partially supported by the National Science Foundation under Grant IIS-0916116 and by the IARPA SCIL program.

## Appendix A. Proofs

The proofs are presented in the order of dependencies. Theorem 3 is the main theorem. The proof of Theorem 2 uses Theorem 3. The proofs of Theorems 4 and 5 follow from Theorem 2. The proof of Lemma 1 follows from Proposition 13.

In the following, unless otherwise noted, $F$ is a rectified first-order sentence, $\mathbf{p}$ is the list of distinct predicate constants $p_1, \dots, p_n$ occurring in $F$, symbols $\mathbf{u}$, $\mathbf{v}$ are lists of distinct predicate variables of the same length as $\mathbf{p}$, and symbols $\mathbf{q}$, $\mathbf{r}$ are lists of distinct predicate names of the same length as $\mathbf{p}$.

### A.1 Proof of Theorem 3

**Theorem 3** *For any rectified sentence $F$, the following sentences are equivalent to each other:*

(*a*) $\mathrm{SM}[F]$;

(*b*) $F \land \forall \mathbf{u}((\mathbf{u} \leq \mathbf{p}) \land \mathit{Nonempty}(\mathbf{u}) \rightarrow \neg \mathit{NSES}_F(\mathbf{u}))$;

(*c*) $F \land \forall \mathbf{u}((\mathbf{u} \leq \mathbf{p}) \land \mathit{Ext\text{-}Loop}_F(\mathbf{u}) \rightarrow \neg \mathit{NSES}_F(\mathbf{u}))$.

The notation that we use in the proof involves *predicate expressions* (Lifschitz, 1994, Section 3.1) of the form

$$\lambda \mathbf{x} F(\mathbf{x}), \tag{45}$$

where $F(\mathbf{x})$ is a formula. If $e$ is (45) and $G(p)$ is a formula containing a predicate constant $p$ of the same arity as the length of $\mathbf{x}$ then $G(e)$ stands for the result of replacing each atomic part of the form $p(\mathbf{t})$ in $G(p)$ with $F(\mathbf{t})$, after renaming the bound variables in $G(p)$ in the usual way, if necessary. For instance, if $G(p)$ is $p(a) \vee p(b)$ then $G(\lambda y(x = y))$ is $x = a \vee x = b$. Substituting a tuple $\mathbf{e}$ of predicate expressions for a tuple $\mathbf{p}$ of predicate constants is defined in a similar way.

**Lemma 5** *Let* $\mathbf{v}$ *be the list of* $\lambda \mathbf{y}^i(p_i(\mathbf{y}^i) \wedge \neg u_i(\mathbf{y}^i))$. *The following formulas are logically valid:*

- $\mathbf{u} \le \mathbf{p} \rightarrow (F^*(\mathbf{u}) \leftrightarrow \mathit{NSES}_F(\mathbf{v}))$;
- $\mathbf{u} \le \mathbf{p} \rightarrow (F^*(\mathbf{v}) \leftrightarrow \mathit{NSES}_F(\mathbf{u}))$.

**Proof**. By induction.

### A.1.1 Proof of Equivalence between (a) and (b) of Theorem 3

It is sufficient to show that

$$\exists \mathbf{u}(\mathbf{u} < \mathbf{p} \wedge F^*(\mathbf{u}))$$

is equivalent to

$$\exists \mathbf{v}(\mathbf{v} \le \mathbf{p} \wedge \mathit{Nonempty}(\mathbf{v}) \wedge \mathit{NSES}_F(\mathbf{v})).$$

*From left to right:* Take $\mathbf{u}$ such that $\mathbf{u} < \mathbf{p} \wedge F^*(\mathbf{u})$. Let $\mathbf{v}$ be the list of $\lambda \mathbf{y}^i(p_i(\mathbf{y}^i) \wedge \neg u_i(\mathbf{y}^i))$.

- Clearly, $\mathbf{v} \le \mathbf{p}$ holds.
- From $\mathbf{u} < \mathbf{p}$, it follows that there are $\mathbf{x}$ and $i$ such that $p_i(\mathbf{x}) \wedge \neg u_i(\mathbf{x})$, from which $\bigvee_i \exists \mathbf{x}^i v_i(\mathbf{x}^i)$ follows, so that $\mathit{Nonempty}(\mathbf{v})$ follows.
- By Lemma 5, $\mathit{NSES}_F(\mathbf{v})$ follows from $\mathbf{u} < \mathbf{p}$ and $F^*(\mathbf{u})$.

*From right to left:* Take $\mathbf{v}$ such that $\mathbf{v} \le \mathbf{p} \wedge \mathit{Nonempty}(\mathbf{v}) \wedge \mathit{NSES}_F(\mathbf{v})$. Let $\mathbf{u}$ be the list of $\lambda \mathbf{y}^i(p_i(\mathbf{y}^i) \wedge \neg v_i(\mathbf{y}^i))$.

- Clearly, $\mathbf{u} \le \mathbf{p}$ holds. Moreover $\neg(\mathbf{u} = \mathbf{p})$ holds. Indeed, if $\mathbf{u} = \mathbf{p}$, then $\forall \mathbf{x}^i \neg v_i(\mathbf{x}^i)$ follows, which contradicts the assumption $\mathit{Nonempty}(\mathbf{v})$. Consequently, $\mathbf{u} < \mathbf{p}$ follows.
- By Lemma 5, $F^*(\mathbf{u})$ follows from $\mathbf{v} \le \mathbf{p}$ and $\mathit{NSES}_F(\mathbf{v})$.

□

### A.1.2 Proof of Proposition 3

**Lemma 6** *Let $I$ be an interpretation of $\sigma$ that contains $\sigma(F)$, and let $\mathbf{q}$, $\mathbf{r}$ be lists of predicate names corresponding to $\mathbf{p}$. Let $Z$ and $Y$ be sets of atoms in the dependency graph of $F$ w.r.t. $I$ such that*

$$p_i(\boldsymbol{\xi}^\diamond) \in Y \text{ iff } I \models q_i(\boldsymbol{\xi}^\diamond)$$

*and*

$$p_i(\boldsymbol{\xi}^\diamond) \in Z \text{ iff } I \models r_i(\boldsymbol{\xi}^\diamond),$$

*where $\boldsymbol{\xi}^\diamond$ is a list of object names. Then*

$$I \models \mathbf{r} \le \mathbf{q} \wedge E_F(\mathbf{r}, \mathbf{q})$$

*iff $Z$ is a subset of $Y$ and there is an edge from an atom in $Z$ to an atom in $Y \setminus Z$ in the dependency graph of $F$ w.r.t. $I$.*

**Proof**. *From left to right:* Assume $I \models \mathbf{r} \le \mathbf{q} \wedge E_F(\mathbf{r}, \mathbf{q})$. The fact that $Z$ is a subset of $Y$ follows from the assumption that $I \models \mathbf{r} \le \mathbf{q}$ and the construction of $Z$ and $Y$. Since

$$I \models \bigvee_{\substack{(p_i(\mathbf{t}), p_j(\mathbf{t}')) \;:\; p_i(\mathbf{t}) \text{ depends on } p_j(\mathbf{t}') \\ \text{in a rule of } F}} \exists \mathbf{z}(r_i(\mathbf{t}) \wedge q_j(\mathbf{t}') \wedge \neg r_j(\mathbf{t}')),$$

where $\mathbf{z}$ is the list of all object variables in $\mathbf{t}$ and $\mathbf{t}'$, there is a substitution $\theta$ that maps object variables in $\mathbf{t}$ and $\mathbf{t}'$ to object names such that

$$I \models \bigvee_{\substack{(p_i(\mathbf{t}), p_j(\mathbf{t}')) \;:\; p_i(\mathbf{t}) \text{ depends on } p_j(\mathbf{t}') \\ \text{in a rule of } F}} r_i(\mathbf{t}\theta) \wedge q_j(\mathbf{t}'\theta) \wedge \neg r_j(\mathbf{t}'\theta).$$

Consequently, there are atoms $p_i(\mathbf{t})$, $p_j(\mathbf{t}')$ such that $p_i(\mathbf{t})$ depends on $p_j(\mathbf{t}')$ in a rule of $F$ and $I \models r_i(\mathbf{t}\theta) \wedge q_j(\mathbf{t}'\theta) \wedge \neg r_j(\mathbf{t}'\theta)$. From $I \models r_i(\mathbf{t}\theta)$ and the construction of $Z$, it follows that $p_i(((\mathbf{t}\theta)^I)^\diamond)$ belongs to $Z$. Also from $I \models q_j(\mathbf{t}'\theta) \wedge \neg r_j(\mathbf{t}'\theta)$, it follows that that $p_j(((\mathbf{t}'\theta)^I)^\diamond)$ belongs to $Y \setminus Z$. Therefore, there is an edge from an atom in $Z$ to an atom in $Y \setminus Z$ in the dependency graph of $F$ w.r.t. $I$.

*From right to left:* Assume that $Z$ is a subset of $Y$ and there is an edge from an atom $p_i(\boldsymbol{\xi}^\diamond)$ in $Z$ to an atom $p_j(\boldsymbol{\eta}^\diamond)$ in $Y \setminus Z$ in the dependency graph of $F$ w.r.t. $I$. Clearly, $I \models \mathbf{r} \le \mathbf{q}$.

From the assumption that $p_i(\boldsymbol{\xi}^\diamond) \in Z$, $p_j(\boldsymbol{\eta}^\diamond) \in Y \setminus Z$ and the construction of $Y$ and $Z$, it follows that $I \models r_i(\boldsymbol{\xi}^\diamond) \wedge q_j(\boldsymbol{\eta}^\diamond) \wedge \neg r_j(\boldsymbol{\eta}^\diamond)$. From the definition of the dependency graph w.r.t. $I$, it follows that there are $p_i(\mathbf{t})$, $p_j(\mathbf{t}')$ such that $p_i(\mathbf{t})$ depends on $p_j(\mathbf{t}')$ in a rule of $F$ with a substitution $\theta$ that maps object variables in $\mathbf{t}$ and $\mathbf{t}'$ to object names such that $(\mathbf{t}\theta)^I = \boldsymbol{\xi}$ and $(\mathbf{t}'\theta)^I = \boldsymbol{\eta}$.

Consequently,

$$I \models \bigvee_{\substack{(p_i(\mathbf{t}), p_j(\mathbf{t}')) \;:\; p_i(\mathbf{t}) \text{ depends on } p_j(\mathbf{t}') \\ \text{in a rule of } F}} r_i(\mathbf{t}\theta) \wedge q_j(\mathbf{t}'\theta) \wedge \neg r_j(\mathbf{t}'\theta),$$

which is equivalent to saying that

$$I \models \bigvee_{\substack{(p_i(\mathbf{t}),p_j(\mathbf{t}')) \ : \ p_i(\mathbf{t}) \text{ depends on } p_j(\mathbf{t}') \\ \text{in a rule of } F}} \exists \mathbf{z}(r_i(\mathbf{t}) \wedge q_j(\mathbf{t}') \wedge \neg r_j(\mathbf{t}')),$$

where $\mathbf{z}$ is the list of all variables in $\mathbf{t}$ and $\mathbf{t}'$. □

**Lemma 7** *A graph $(V, E)$ is strongly connected iff, for any nonempty proper subset $U$ of $V$, there is an edge from $U$ to $V \setminus U$.*

**Proof**. Follows from the definition of a strongly connected graph. □

**Proposition 3** *Let $\mathbf{q}$ be a list of predicate names corresponding to $\mathbf{p}$, and let $Y$ be a set of atoms in the dependency graph of $F$ w.r.t. $I$ such that*

$$p_i(\boldsymbol{\xi}^\diamond) \in Y \text{ iff } I \models q_i(\boldsymbol{\xi}^\diamond),$$

*where $\boldsymbol{\xi}^\diamond$ is a list of object names. Then $I \models Loop_F(\mathbf{q})$ iff $Y$ is a loop of $F$ w.r.t. $I$.*

**Proof**. *From left to right:* Assume that $I \models Loop_F(\mathbf{q})$. From $I \models Nonempty(\mathbf{q})$, it follows that $Y$ is nonempty.

Take any nonempty proper subset $Z$ of $Y$. Let $\mathbf{r}$ be the list of predicate names such that

$$I \models r_i(\boldsymbol{\xi}^\diamond) \text{ iff } p_i(\boldsymbol{\xi}^\diamond) \in Z.$$

It is clear that

$$I \models Nonempty(\mathbf{r}) \wedge \mathbf{r} < \mathbf{q}.$$

Consequently, from $I \models Loop_F(\mathbf{q})$, it follows that $I \models E_F(\mathbf{r}, \mathbf{q})$. By Lemma 6, there is an edge from an atom in $Z$ to an atom in $Y \setminus Z$. Consequently, by Lemma 7, $Y$ induces a strongly connected subgraph and thus a loop of $F$ w.r.t. $I$.

*From right to left:* Let $Y$ be loop of $F$ w.r.t. $I$ and $\mathbf{q}$ a list of predicate names such that

$$I \models q_i(\boldsymbol{\xi}^\diamond) \text{ iff } p_i(\boldsymbol{\xi}^\diamond) \in Y.$$

Since $Y$ is nonempty, $I \models Nonempty(\mathbf{q})$.

Consider any list of predicate names $\mathbf{r}$ such that

$$I \models Nonempty(\mathbf{r}) \wedge \mathbf{r} < \mathbf{q}.$$

Let $Z$ be a set of vertices in the dependency graph of $F$ w.r.t. $I$ such that

$$p_i(\boldsymbol{\xi}^\diamond) \in Z \text{ iff } I \models r_i(\boldsymbol{\xi}^\diamond).$$

Clearly, $Z$ is a nonempty proper subset of $Y$. Since $Y$ induces a strongly connected subgraph, by Lemma 7, there is an edge from an atom in $Z$ to an atom in $Y \setminus Z$. Consequently by Lemma 6, $I \models E_F(\mathbf{r}, \mathbf{q})$. □

### A.1.3 Proof of Proposition 4

**Proposition 4** *Let* $\mathbf{q}$ *be a list of predicate names corresponding to* $\mathbf{p}$*, and let* $Y$ *be a set of atoms in the dependency graph of* $F$ *w.r.t.* $I$ *such that*

$$p_i(\boldsymbol{\xi}^\diamond) \in Y \text{ iff } I \models q_i(\boldsymbol{\xi}^\diamond),$$

*where* $\boldsymbol{\xi}^\diamond$ *is a list of object names. Then*

$$I \models \mathit{Nonempty}(\mathbf{q}) \wedge \forall \mathbf{v}((\mathbf{v} \leq \mathbf{q}) \wedge \mathit{Loop}_F(\mathbf{v}) \rightarrow E_F(\mathbf{v}, \mathbf{q}))$$

*iff* $Y$ *is an unbounded set of* $F$ *w.r.t.* $I$*.*

**Proof**. *From left to right:* Assume

$$I \models \mathit{Nonempty}(\mathbf{q}) \wedge \forall \mathbf{v}(\mathbf{v} \leq \mathbf{q} \wedge \mathit{Loop}_F(\mathbf{v}) \rightarrow E_F(\mathbf{v}, \mathbf{q})). \tag{46}$$

Since $I \models \mathit{Nonempty}(\mathbf{q})$, it is clear that $Y$ is nonempty.

Take any subset $Z$ of $Y$ that is a loop of $F$ w.r.t. $I$. Let $\mathbf{r}$ be a list of predicate names such that

$$I \models r_i(\boldsymbol{\xi}^\diamond) \text{ iff } p_i(\boldsymbol{\xi}^\diamond) \in Z.$$

Since $Z$ is a subset of $Y$, it is clear that $I \models \mathbf{r} \leq \mathbf{q}$. Since $Z$ is a loop of $F$ w.r.t. $I$, by Proposition 3, $I \models \mathit{Loop}_F(\mathbf{r})$. Consequently, from (46) it follows that $I \models E_F(\mathbf{r}, \mathbf{q})$. By Lemma 6, there is an edge from an atom in $Z$ to an atom in $Y \setminus Z$. Therefore, $Y$ is an unbounded set of $F$ w.r.t. $I$.

*From right to left:* Let $Y$ be an unbounded set of $F$ w.r.t. $I$. Since $Y$ is nonempty, it is clear that $I \models \mathit{Nonempty}(\mathbf{q})$.

Take any list of predicate names $\mathbf{r}$ such that $I \models \mathbf{r} \leq \mathbf{q} \wedge \mathit{Loop}_F(\mathbf{r})$. Let $Z$ be a set of vertices in the dependency graph of $F$ w.r.t. $I$ such that

$$p_i(\boldsymbol{\xi}^\diamond) \in Z \text{ iff } I \models r_i(\boldsymbol{\xi}^\diamond).$$

By Proposition 3, $Z$ is a loop of $F$ w.r.t. $I$. It is clear that $Z$ is a subset of $Y$. Since $Y$ is an unbounded set of $F$ w.r.t. $I$, there is an edge from $Z$ to $Y \setminus Z$. Consequently by Lemma 6, $I \models E_F(\mathbf{r}, \mathbf{q})$. □

### A.1.4 Proof of Proposition 5

**Proposition 5** *For any negative formula* $F$*, formula*

$$\mathit{NSES}_F(\mathbf{u}) \leftrightarrow F$$

*is logically valid.*

**Proof**. The proof follows immediately from the following two lemmas, which can be proved by induction. □

**Lemma 8** *For any formula $F$,*

$$NSES_F(\mathbf{u}) \rightarrow F$$

*is logically valid.*

**Lemma 9** *Let $F$ be a formula, and let $S_F$ be the set of $p_i(\mathbf{t})$ that has a strictly positive occurrence in $F$. Formula*

$$F \wedge \bigwedge_{p_i(\mathbf{t}) \in S_F} \forall \mathbf{z} \neg v_i(\mathbf{t}) \rightarrow NSES_F(\mathbf{v}) \tag{47}$$

*is logically valid, where $\mathbf{z}$ is the tuple of variables in $\mathbf{t}$ that are not free in $F$.*

#### A.1.5 Proof of Equivalence between (b) and (c) of Theorem 3

**Lemma 10** *Let $F$ be a rectified formula, let $S_F^+$ be the set of all atoms $p_i(\mathbf{t})$ that have a positive occurrence in $F$ that does not belong to a negative formula, and let $S_F^-$ be the set of all atoms $p_i(\mathbf{t})$ that have a negative occurrence in $F$ that does not belong to a negative formula.*[16] *The following formulas are logically valid, where $\mathbf{z}$ is the list of all variables in $\mathbf{t}$ that are not free in $F$.*

*(a)* $(\mathbf{v} \leq \mathbf{u}) \wedge \bigwedge_{p_i(\mathbf{t}) \in S_F^+} \forall \mathbf{z}(u_i(\mathbf{t}) \rightarrow v_i(\mathbf{t})) \wedge NSES_F(\mathbf{v}) \rightarrow NSES_F(\mathbf{u})$;

*(b)* $(\mathbf{v} \leq \mathbf{u}) \wedge \bigwedge_{p_i(\mathbf{t}) \in S_F^-} \forall \mathbf{z}(u_i(\mathbf{t}) \rightarrow v_i(\mathbf{t})) \wedge NSES_F(\mathbf{u}) \rightarrow NSES_F(\mathbf{v})$.

**Proof**. Both parts are proved simultaneously by induction on $F$.

*Case 1:* $F$ is an atom $p_i(\mathbf{t})$.

Part (a): $NSES_F(\mathbf{v})$ entails $NSES_F(\mathbf{u})$ under the assumption

$$\bigwedge_{p_i(\mathbf{t}) \in S_F^+} \forall \mathbf{z}(u_i(\mathbf{t}) \rightarrow v_i(\mathbf{t})).$$

Part (b): $NSES_F(\mathbf{u})$ entails $NSES_F(\mathbf{v})$ under the assumption $\mathbf{v} \leq \mathbf{u}$.

*Case 2:* $F$ is $\bot$ or an equality. It is clear since $NSES_F(\mathbf{v})$ and $NSES_F(\mathbf{u})$ are the same as $F$.

*Case 3:* $F$ is $G \wedge H$ or $G \vee H$. Follows from I.H.

*Case 4:* $F$ is $G \rightarrow H$.

Part (a): Assume

$$(\mathbf{v} \leq \mathbf{u}) \wedge \bigwedge_{p_i(\mathbf{t}) \in S_F^+} \forall \mathbf{z}(u_i(\mathbf{t}) \rightarrow v_i(\mathbf{t})). \tag{48}$$

We need to show that

$$(NSES_G(\mathbf{v}) \rightarrow NSES_H(\mathbf{v})) \wedge (G \rightarrow H)$$

16. Note that we distinguish between formula being negative and an occurrence being negative. See at the end of Section 2.

entails

$$(NSES_G(\mathbf{u}) \rightarrow NSES_H(\mathbf{u})) \wedge (G \rightarrow H).$$

Note that

$$\bigwedge_{p_i(\mathbf{t}) \in S_G^-} \forall \mathbf{z}(u_i(\mathbf{t}) \rightarrow v_i(\mathbf{t}))$$

and

$$\bigwedge_{p_i(\mathbf{t}) \in S_H^+} \forall \mathbf{z}(u_i(\mathbf{t}) \rightarrow v_i(\mathbf{t}))$$

are entailed by formula (48). By I.H., $NSES_G(\mathbf{u})$ entails $NSES_G(\mathbf{v})$ and $NSES_H(\mathbf{v})$ entails $NSES_H(\mathbf{u})$.

Part (b): Similar to Part (a).

*Case 5:* $F$ is $\forall x\, G$

Part (a): Assume

$$(\mathbf{v} \leq \mathbf{u}) \wedge \bigwedge_{p_i(\mathbf{t}) \in S_F^+} \forall \mathbf{z}(u_i(\mathbf{t}) \rightarrow v_i(\mathbf{t})) \wedge \forall x NSES_G(\mathbf{v}).$$

From the assumption $NSES_G(\mathbf{v})$, $G$ follows by Lemma 8. Also

$$\bigwedge_{p_i(\mathbf{t}) \in S_G^+} \forall \mathbf{z}'(u_i(\mathbf{t}) \rightarrow v_i(\mathbf{t}))$$

follows, where $\mathbf{z}'$ is the list of all variables in $\mathbf{t}$ that are not free in $G$, so that by I.H. on $G$, $NSES_G(\mathbf{u})$ holds from the assumption. Since $x$ is not free in the assumption, $\forall x NSES_G(\mathbf{u})$ holds as well.

Part (b): Similar to Part (a).

*Case 6:* $F$ is $\exists x\ G$.

Part (a): Assume

$$(\mathbf{v} \leq \mathbf{u}) \wedge \bigwedge_{p_i(\mathbf{t}) \in S_F^+} \forall \mathbf{z}(u_i(\mathbf{t}) \rightarrow v_i(\mathbf{t})) \wedge \exists x NSES_G(\mathbf{v}). \tag{49}$$

Take $x$ such that

$$(\mathbf{v} \leq \mathbf{u}) \wedge \bigwedge_{p_i(\mathbf{t}) \in S_F^+} \forall \mathbf{z}(u_i(\mathbf{t}) \rightarrow v_i(\mathbf{t})) \wedge NSES_G(\mathbf{v}). \tag{50}$$

From $NSES_G(\mathbf{v})$, by Lemma 8, $G$ follows. Also

$$\bigwedge_{p_i(\mathbf{t}) \in S_G^+} \forall \mathbf{z}'(u_i(\mathbf{t}) \rightarrow v_i(\mathbf{t}))$$

follows, where $\mathbf{z}'$ is the list of all variables in $\mathbf{t}$ that are not free in $G$. By I.H. on $G$, $NSES_G(\mathbf{u})$ holds under the assumption (50). Consequently, $\exists x NSES_G(\mathbf{u})$ holds from the same assumption. Since $x$ is not free in (49), we conclude that $\exists x NSES_G(\mathbf{u})$ holds from the assumption (49).

Part (b): Similar to Part (a). $\square$

**Lemma 11** *For any rectified formula $F$,*

$$(\mathbf{v} \leq \mathbf{u}) \wedge \neg E_F(\mathbf{v}, \mathbf{u}) \wedge NSES_F(\mathbf{u}) \rightarrow NSES_F(\mathbf{v})$$

*is logically valid.*

**Proof**. By induction on $F$.

*Case 1:* $F$ is an atom $p_i(\mathbf{t})$. $NSES_F(\mathbf{u})$ entails $NSES_F(\mathbf{v})$ under the assumption $\mathbf{v} \leq \mathbf{u}$.

*Case 2:* $F$ is $\bot$ or equality. It is clear since $NSES_F(\mathbf{v})$ and $NSES_F(\mathbf{u})$ are the same as $F$.

*Case 3:* $F$ is $G \wedge H$ or $G \vee H$. Follows from I.H.

*Case 4:* $F$ is $G \rightarrow H$. Assume

$$(\mathbf{v} \leq \mathbf{u}) \wedge \neg E_F(\mathbf{v}, \mathbf{u}) \wedge NSES_F(\mathbf{u})$$

and $NSES_G(\mathbf{v})$. From $NSES_F(\mathbf{u})$, by Lemma 8, we conclude $G \rightarrow H$. From $NSES_G(\mathbf{v})$, by Lemma 8, $G$ follows, and consequently $H$.

Assume $\neg NSES_H(\mathbf{v})$ for the sake of contradiction. By Lemma 9, from $H$ and $\neg NSES_H(\mathbf{v})$, it follows that

$$\bigvee_{p_i(\mathbf{t})\ :\ p_i(\mathbf{t}) \text{ occurs strictly positively in } H} \exists \mathbf{x} v_i(\mathbf{t}) \tag{51}$$

, where $\mathbf{x}$ is the list of variables in $\mathbf{t}$ that are not free in $H$.

Since $F$ is rectified, the variables in $F$ can be partitioned into three sets: the list of variables $\mathbf{x}$ that are not free in $H$, the list of variables $\mathbf{y}$ that are not free in $G$, and the rest. Note that $\neg E_F(\mathbf{v}, \mathbf{u})$ entails

$$\bigwedge_{\substack{(p_i(\mathbf{t}), p_j(\mathbf{t}'))\ :\ p_i(\mathbf{t}) \text{ depends on } p_j(\mathbf{t}') \text{ in a rule } G \rightarrow H \text{ in } F \\ p_i(\mathbf{t}) \text{ occurs in } H, p_j(\mathbf{t}') \text{ occurs in } G}} \Big(\exists \mathbf{x} v_i(\mathbf{t}) \rightarrow \forall \mathbf{y}(u_j(\mathbf{t}') \rightarrow v_j(\mathbf{t}'))\Big), \tag{52}$$

where $\mathbf{x}$ is the list of all variables in $\mathbf{t}$ that are not free in $H$, and $\mathbf{y}$ is the list of all variables in $\mathbf{t}'$ that are not free in $G$. From (51) and (52), we conclude

$$\bigwedge_{p_j(\mathbf{t}')\ :\ p_j(\mathbf{t}') \text{ occurs positively and not in a negative subformula of } G} \forall \mathbf{y}(u_j(\mathbf{t}') \rightarrow v_j(\mathbf{t}')).$$

From this, together with the assumption $(\mathbf{v} \leq \mathbf{u})$ and $NSES_G(\mathbf{v})$, by Lemma 10 (a), $NSES_G(\mathbf{u})$ follows. Thus $NSES_H(\mathbf{u})$ follows from $NSES_F(\mathbf{u})$ and $NSES_G(\mathbf{u})$. Since $\neg E_F(\mathbf{v}, \mathbf{u})$ entails $\neg E_H(\mathbf{v}, \mathbf{u})$, by I.H. on $H$, $NSES_H(\mathbf{v})$ follows, which contradicts the assumption.

*Case 5:* $F$ is $\forall x G$ or $\exists x G$. Follows from I.H. $\square$

**Lemma 12**

$$Nonempty(\mathbf{u}) \rightarrow \exists \mathbf{v}(\mathbf{v} \leq \mathbf{u} \wedge \textit{Ext-Loop}_F(\mathbf{v}) \wedge \neg E_F(\mathbf{v}, \mathbf{u}))$$

*is logically valid.*

**Proof**. Take any list $\mathbf{q}$ of predicate names, and any interpretation $I$ that satisfies $Nonempty(\mathbf{q})$. Let $Y$ be a set of vertices in the dependency graph of $F$ w.r.t. $I$ such that

$$p_i(\boldsymbol{\xi}^\diamond) \in Y \text{ iff } I \models q_i(\boldsymbol{\xi}^\diamond).$$

Consider the subgraph $G$ of the dependency graph of $F$ w.r.t. $I$ that is induced by $Y$. If $Y$ is an unbounded set w.r.t. $I$, by Proposition 4, $I \models \textit{Ext-Loop}_F(\mathbf{q})$. So

$$I \models \mathbf{q} \leq \mathbf{q} \wedge \textit{Ext-Loop}_F(\mathbf{q}) \wedge \neg E_F(\mathbf{q}, \mathbf{q}).$$

Otherwise, consider the graph $G'$ that is obtained from $G$ by collapsing strongly connected components of $G$, i.e., the vertices of $G'$ are the strongly connected components of $G$ and $G'$ has an edge from $V$ to $V'$ if $G$ has an edge from a vertex in $V$ to a vertex in $V'$. Since we assumed that $Y$ is not an unbounded set w.r.t. $I$, there exists a vertex $Z$ in $G'$ that has no outgoing edges. Consider the list of predicate names $\mathbf{r}$ such that

$$I \models r_i(\boldsymbol{\xi}^\diamond) \text{ iff } p_i(\boldsymbol{\xi}^\diamond) \in Z.$$

It is clear that $I \models \mathbf{r} \leq \mathbf{q}$. By Proposition 3, $I \models \textit{Loop}_F(\mathbf{r})$ thus $I \models \textit{Ext-Loop}_F(\mathbf{r})$. Since there is no edge from $Z$ to $Y \setminus Z$, by Lemma 6, $I \models \neg E_F(\mathbf{r}, \mathbf{q})$. Consequently, the claim follows. □

**Proof of Equivalence Between (b) and (c) of Theorem 3**

*From (b) to (c):* Clear from that the formula $\textit{Ext-Loop}_F(\mathbf{u}) \rightarrow Nonempty(\mathbf{u})$ is logically valid.

*From (c) to (b):* Assume

$$F \wedge \forall \mathbf{v}(\mathbf{v} \leq \mathbf{p} \wedge \textit{Ext-Loop}_F(\mathbf{v}) \rightarrow \neg \textit{NSES}_F(\mathbf{v})).$$

Take any $\mathbf{u}$ such that $\mathbf{u} \leq \mathbf{p} \wedge Nonempty(\mathbf{u})$. By Lemma 12, it follows from $Nonempty(\mathbf{u})$ that there exists $\mathbf{v}$ such that $\mathbf{v} \leq \mathbf{u} \wedge \textit{Ext-Loop}_F(\mathbf{v}) \wedge \neg E_F(\mathbf{v}, \mathbf{u})$. It is clear that $\mathbf{v} \leq \mathbf{p}$ follows from $\mathbf{v} \leq \mathbf{u}$ and $\mathbf{u} \leq \mathbf{p}$. It follows from the assumption that $\neg \textit{NSES}_F(\mathbf{v})$. Then by Lemma 11, $\neg \textit{NSES}_F(\mathbf{u})$ follows from $\mathbf{v} \leq \mathbf{u}$ and $\neg E_F(\mathbf{v}, \mathbf{u})$. □

### A.2 Proof of Theorem 2

**Lemma 3** *Let $F$ be a rectified sentence of a signature $\sigma$ (possibly containing function constants of positive arity), and let $I$ be an interpretation of $\sigma$ that satisfies $F$. If $F$ is bounded w.r.t. $I$,*

$$I \models \exists \mathbf{u}(\mathbf{u} \leq \mathbf{p} \wedge \textit{Ext-Loop}_F(\mathbf{u}) \wedge \textit{NSES}_F(\mathbf{u}))$$

*iff there is a finite loop $Y$ of $F$ w.r.t. $I$ such that*

$$I \models \Big(\bigwedge Y \wedge NES_F(Y)\Big).$$

**Proof**. *From left to right:* Assume

$$I \models \mathbf{q} \leq \mathbf{p} \wedge \textit{Ext-Loop}_F(\mathbf{q}) \wedge NSES_F(\mathbf{q})$$

for some list of predicate names $\mathbf{q}$. Consider $Y$ to be the set of vertices in the dependency graph of $F$ w.r.t. $I$ such that

$$p_i(\boldsymbol{\xi}^\diamond) \in Y \text{ iff } I \models q_i(\boldsymbol{\xi}^\diamond).$$

Since $I \models \textit{Ext-Loop}_F(\mathbf{q})$, by Proposition 3 and Proposition 4, it follows that $Y$ is an extended loop of $F$ w.r.t. $I$. Since $I \models q_i(\boldsymbol{\xi}^\diamond)$ for all $p_i(\boldsymbol{\xi}^\diamond) \in Y$ and $I \models \mathbf{q} \leq \mathbf{p}$, it follows that $I$ satisfies every atom in $Y$. Together with the assumption that $F$ is bounded w.r.t. $I$, this implies that set $Y$ is finite. Since $I \models NSES_F(\mathbf{q})$ and $Y$ is finite, by Lemma 2, it follows that $I \models NES_F(Y)$.

*From right to left:* Consider any finite loop $Y$ of $F$ w.r.t. $I$. Assume

$$I \models \bigwedge Y \wedge NES_F(Y).$$

Let $\mathbf{q}$ be a list of predicate names such that

$$I \models q_i(\boldsymbol{\xi}^\diamond) \text{ iff } p_i(\boldsymbol{\xi}^\diamond) \in Y.$$

- $I \models \mathbf{q} \leq \mathbf{p}$ follows from the construction of $\mathbf{q}$ and $I \models \bigwedge Y$.
- Since $Y$ is a loop of $F$ w.r.t. $I$, by Proposition 3, $I \models \textit{Loop}_F(\mathbf{q})$, and consequently, $I \models \textit{Ext-Loop}_F(\mathbf{q})$.
- From $I \models NES_F(Y)$, by Lemma 2, $I \models NSES_F(\mathbf{q})$.

Consequently, $I \models \exists \mathbf{u}(\mathbf{u} \leq \mathbf{p} \wedge \textit{Ext-Loop}_F(\mathbf{u}) \wedge NSES_F(\mathbf{u}))$. □

**Theorem 2** *Let $F$ be a rectified sentence of a signature $\sigma$ (possibly containing function constants of positive arity), and let $I$ be an interpretation of $\sigma$ that satisfies $F$. If $F$ is bounded w.r.t. $I$, then the following conditions are equivalent to each other:*

*(a) $I$ satisfies* SM[$F$]*;*

*(b) for every nonempty finite set $Y$ of atoms formed from predicate constants in $\sigma(F)$ and object names for $|I|$, $I$ satisfies $LF_F(Y)$;*

*(c) for every finite loop $Y$ of $F$ w.r.t. $I$, $I$ satisfies $LF_F(Y)$.*

**Proof**. *Between (a) and (c):* By Theorem 3 and Lemma 3.

*Between (b) and (c):*

- *From (b) to (c):* Clear.
- *From (c) to (b):* Assume that $I$ satisfies $LF_F(L)$ for every finite loop $L$ of $F$ w.r.t $I$. Consider any nonempty finite set $Y$ of atoms formed from predicate constants in $\sigma(F)$ and object names such that $I \models \bigwedge Y$. Let $\mathbf{q}$ be a list of predicate names such that

$$I \models q_i(\boldsymbol{\xi}^\diamond) \text{ iff } p_i(\boldsymbol{\xi}^\diamond) \in Y.$$

Since $Y$ is nonempty, it is clear that $Nonempty(\mathbf{q})$ follows. In view of Lemma 12, there is a list of predicate names $\mathbf{r}$ such that

$$I \models \mathbf{r} \leq \mathbf{q} \wedge \textit{Ext-Loop}_F(\mathbf{r}) \wedge \neg E_F(\mathbf{r}, \mathbf{q}). \tag{53}$$

Consider $Z$ to be the set of vertices in the dependency graph of $F$ w.r.t. $I$ such that

$$p_i(\boldsymbol{\xi}^\diamond) \in Z \text{ iff } I \models r_i(\boldsymbol{\xi}^\diamond).$$

Since $I \models \textit{Ext-Loop}_F(\mathbf{r})$, by Proposition 3 and Proposition 4, $Z$ is an extended loop of $F$ w.r.t. $I$. Clearly, $I \models \bigwedge Z$ since $Z \subseteq Y$ and $I \models \bigwedge Y$. Since $F$ is bounded w.r.t. $I$, and $Z$ is satisfied by $I$, it follows that $Z$ is a finite loop of $F$ w.r.t. $I$. Since $I \models \mathbf{r} \leq \mathbf{q} \wedge \neg E_F(\mathbf{r}, \mathbf{q})$, $Z$ is a subset of $Y$ and, by Lemma 6, there is no edge from $Z$ to $Y \setminus Z$ in the dependency graph of $F$ w.r.t. $I$. Since $I \models LF_F(Z)$, we conclude that $I \models \neg NES_F(Z)$, and by Lemma 2, $I \models \neg NSES_F(\mathbf{r})$. From (53) and that $I \models \neg NSES_F(\mathbf{r})$, by Lemma 11, we have $I \models \neg NSES_F(\mathbf{q})$. By Lemma 2 again, $I \models \neg NES_F(Y)$. Consequently, $I \models LF_F(Y)$.

□

### A.3 Proof of Proposition 6

**Proposition 6** *If a rectified formula $F$ of a signature $\sigma$ is bounded, then $F$ is bounded w.r.t. any interpretation of $\sigma$ that satisfies $\mathrm{CET}_\sigma$.*

**Lemma 13** *For any terms $t_1$ and $t_2$ of signature $\sigma$, any interpretation $I$ that satisfies $\mathrm{CET}_\sigma$, and any substitution $\theta$ from object variables in $t_1$ and $t_2$ to object names such that $(t_1\theta)^I = (t_2\theta)^I$, Robinson's unification algorithm (Robinson, 1965), when applied to $t_1$ and $t_2$, returns a most general unifier (mgu) $\gamma$ of $t_1$ and $t_2$ such that*

*(a) $t_1\gamma = t_2\gamma$, and*

*(b) for every variable $x$ in $t_1$ or $t_2$, $(x\gamma\theta)^I = (x\theta)^I$.*

**Proof**. From the assumptions, by Lemma 5.1 from the work of Kunen (1987), $t_1$ and $t_2$ are unifiable, in which case Robinson's algorithm returns a mgu for $t_1$ and $t_2$ that maps variables occurring in $t_1$ and $t_2$ into terms. Given this, part (b) can be proven by induction. □

The proof of Proposition 6 follows from the following lemma.

**Lemma 14** *Let $F$ be a rectified sentence of a signature $\sigma$, and let $I$ be an interpretation of $\sigma$ that satisfies* $\text{CET}_\sigma$*. For any path*

$$\langle p_1(\boldsymbol{\xi}_1^\diamond), p_2(\boldsymbol{\xi}_2^\diamond), \dots, p_k(\boldsymbol{\xi}_k^\diamond), p_{k+1}(\boldsymbol{\xi}_{k+1}^\diamond)\rangle \tag{54}$$

*in the dependency graph of $F$ w.r.t $I$, there is a path*

$$\langle p_1(\mathbf{u}_1), p_2(\mathbf{u}_2), \dots, p_k(\mathbf{u}_k), p_{k+1}(\mathbf{u}_{k+1})\rangle$$

*in the first-order dependency graph of $F$ with a substitution $\theta$ that maps object variables in $\mathbf{u}_i$ to object names such that $(\mathbf{u}_i\theta)^I = \boldsymbol{\xi}_i$ for all $i$.*

**Proof**. Each edge $(p_i(\boldsymbol{\xi}_i^\diamond), p_{i+1}(\boldsymbol{\xi}_{i+1}^\diamond))$ in (54) is obtained from a pair of atoms $(p_i(\mathbf{t}_i), p_{i+1}(\mathbf{t}'_i))$ and a substitution $\theta_i$ such that $p_i(\mathbf{t}_i)$ depends on $p_{i+1}(\mathbf{t}'_i)$ in a rule of $F$, and

$$(\mathbf{t}_1\theta_1)^I = \boldsymbol{\xi}_1, \quad (\mathbf{t}'_i\theta_i)^I = (\mathbf{t}_{i+1}\theta_{i+1})^I = \boldsymbol{\xi}_{i+1} (1 \le i < k), \quad (\mathbf{t}'_k\theta_k)^I = \boldsymbol{\xi}_{k+1}. \tag{55}$$

For simplicity we assume that each pair $(p_i(\mathbf{t}_i), p_{i+1}(\mathbf{t}'_i))$ considered above has no common variables with another pair by first renaming variables. This allows us to use one substitution $\theta = \theta_1 \dots \theta_k$ in place of individual $\theta_i$ in the rest of the proof.

We will show by induction that, for each $j$ where $j \in \{1 \dots k\}$, there are substitutions $\sigma_i^j$ $(1 \le i \le j)$ from variables in $\mathbf{t}_i$ and $\mathbf{t}'_i$ to terms such that

(a) $\langle p_1(\mathbf{t}_1)\sigma_1^j, p_2(\mathbf{t}_2)\sigma_2^j, \dots, p_j(\mathbf{t}_j)\sigma_j^j, p_{j+1}(\mathbf{t}'_j)\sigma_j^j\rangle$ is a path in the first-order dependency graph of $F$, and

(b) $(\mathbf{t}_i\sigma_i^j\theta)^I = \boldsymbol{\xi}_i$ for all $1 \le i \le j$, and $(\mathbf{t}'_j\sigma_j^j\theta)^I = \boldsymbol{\xi}_{j+1}$.

When $j = 1$, we take $\sigma_i^j$ to be an identity substitution. Clearly, conditions (a) and (b) are satisfied.

Otherwise, by I.H. we assume that, for some $j$ in $\{1, \dots, k-1\}$, there are substitutions $\sigma_1^j, \dots, \sigma_j^j$ such that conditions (a) and (b) are satisfied. We will prove that there are substitutions $\sigma_i^{j+1}$ $(1 \le i \le j+1)$ from variables in $\mathbf{t}_i$ and $\mathbf{t}'_i$ to terms such that

(a') $\langle p_1(\mathbf{t}_1)\sigma_1^{j+1}, p_2(\mathbf{t}_2)\sigma_2^{j+1}, \dots, p_{j+1}(\mathbf{t}_{j+1})\sigma_{j+1}^{j+1}, p_{j+2}(\mathbf{t}'_{j+1})\sigma_{j+1}^{j+1}\rangle$ is a path in the first-order dependency graph of $F$, and

(b') $(\mathbf{t}_i\sigma_i^{j+1}\theta)^I = \boldsymbol{\xi}_i$ for all $1 \le i \le j+1$, and $(\mathbf{t}'_{j+1}\sigma_{j+1}^{j+1}\theta)^I = \boldsymbol{\xi}_{j+2}$.

From I.H., we have $(\mathbf{t}'_j\sigma_j^j\theta)^I = \boldsymbol{\xi}_{j+1}$ and from (55) we have $(\mathbf{t}_{j+1}\theta)^I = \boldsymbol{\xi}_{j+1}$. By Lemma 13 there is a substitution $\gamma$ from variables in $\mathbf{t}'_j\sigma_j^j$ or $\mathbf{t}_{j+1}$ to terms such that $\mathbf{t}'_j\sigma_j^j\gamma = \mathbf{t}_{j+1}\gamma$ and for any variable $x$ in $\mathbf{t}'_j\sigma_j^j$ or $\mathbf{t}_{j+1}$,

$$(x\gamma\theta)^I = (x\theta)^I. \tag{56}$$

We define $\sigma_i^{j+1}$ as

- $\sigma_i^j\gamma$ when $1 \le i \le j$ and

- $\gamma$ when $i = j+1$.

It is easy to check that condition (a') is satisfied. To check that condition (b') is satisfied, consider any variable $x$ in the set

$$\{\mathbf{t}_1\sigma_1^j, \mathbf{t}_2\sigma_2^j, \ldots, \mathbf{t}_j\sigma_j^j, \mathbf{t}'_j\sigma_j^j, \mathbf{t}_{j+1}, \mathbf{t}'_{j+1}\}. \tag{57}$$

If $x$ is in $\mathbf{t}'_j\sigma_j^j$ or $\mathbf{t}_{j+1}$, by (56), $(x\gamma\theta)^I = (x\theta)^I$. Otherwise, since $\gamma$ does not change the variables that are not in $\mathbf{t}'_j\sigma_j^j$ or $\mathbf{t}_{j+1}$, $(x\gamma\theta)^I = (x\theta)^I$. Consequently, for any variable $x$ in (57), we get $(x\gamma\theta)^I = (x\theta)^I$. It remains to check the following.

- For $1 \le i \le j$, $(\mathbf{t}_i\sigma_i^{j+1}\theta)^I = (\mathbf{t}_i\sigma_i^j\gamma\theta)^I = (\mathbf{t}_i\sigma_i^j\theta)^I$. The last one is equal to $\boldsymbol{\xi}_i$ by I.H.
- $(\mathbf{t}_{j+1}\sigma_{j+1}^{j+1}\theta)^I = (\mathbf{t}_{j+1}\gamma\theta)^I = (\mathbf{t}_{j+1}\theta)^I$. The last one is equal to $\boldsymbol{\xi}_{j+1}$ by (55).
- $(\mathbf{t}'_{j+1}\sigma_{j+1}^{j+1}\theta)^I = (\mathbf{t}'_{j+1}\gamma\theta)^I = (\mathbf{t}'_{j+1}\theta)^I$. The last one is equal to $\boldsymbol{\xi}_{j+2}$ by (55).

□

### A.4 Proof of Proposition 7

**Proposition 7** *For any rectified sentence $F$ of a signature $\sigma$ and for any interpretation $I$ of $\sigma$ that satisfies* $\mathrm{CET}_\sigma$*, $I$ is a model of*

$$\{LF_F(Y) \mid Y \text{ is a finite first-order loop of } F\}$$

*iff $I$ is a model of*

$$\{LF_F(Y) \mid Y \text{ is a finite loop of } F \text{ w.r.t. } I\}.$$

The proof follows immediately from the following fact and Lemma 15.

**Fact 1** *Let $F$ be a rectified sentence of a signature $\sigma$, and let $I$ be an interpretation of $\sigma$. For any first-order loop $Y$ of $F$ and any substitution $\theta$ that maps variables in $Y$ to object names, $Y' = \{p_i(\boldsymbol{\xi}^\diamond) \mid p_i(\mathbf{t}) \in Y\theta,\ \mathbf{t}^I = \boldsymbol{\xi}\}$ is a loop of $F$ w.r.t. $I$.*

**Lemma 15** *Let $F$ be a rectified sentence of a signature $\sigma$, and let $I$ be an interpretation of $\sigma$. If $I$ satisfies* $\mathrm{CET}_\sigma$*, then, for any finite loop $Y'$ of $F$ w.r.t. $I$, there is a finite loop $Y$ of $F$ with a substitution $\theta$ that maps variables in $Y$ to object names such that $Y' = \{p_i(\boldsymbol{\xi}^\diamond) \mid p_i(\mathbf{t}) \in Y,\ (\mathbf{t}\theta)^I = \boldsymbol{\xi}\}$.*

**Proof**. Without loss of generality, consider a path

$$\langle p_1(\boldsymbol{\xi}_1^\diamond), p_2(\boldsymbol{\xi}_2^\diamond), \ldots, p_k(\boldsymbol{\xi}_k^\diamond), p_1(\boldsymbol{\xi}_1^\diamond)\rangle$$

($k \ge 1$) in the dependency graph of $F$ w.r.t. $I$ that consists of the vertices in $Y'$. Since $I \models \mathrm{CET}_\sigma$, by Lemma 14, there is a path

$$\langle p_1(\mathbf{u}_1), p_2(\mathbf{u}_2), \ldots, p_k(\mathbf{u}_k), p_1(\mathbf{u}_{k+1})\rangle$$

in the first-order dependency graph of $F$ with a substitution $\theta$ that maps variables in $\mathbf{u}_i$ to object names such that $(\mathbf{u}_i\theta)^I = \boldsymbol{\xi}_i$ for all $1 \leq i \leq k$, and $(\mathbf{u}_{k+1}\theta)^I = \boldsymbol{\xi}_1$. Since $(\mathbf{u}_{k+1}\theta)^I = (\mathbf{u}_1\theta)^I$, by Lemma 13, there is a unifier $\gamma$ for $\mathbf{u}_{k+1}$ and $\mathbf{u}_1$ such that, for any variable $x$ in $\mathbf{u}_{k+1}$ or $\mathbf{u}_1$, $(x\gamma\theta)^I = (x\theta)^I$. Consequently,

$$\{p_1(\mathbf{u}_1\gamma), p_2(\mathbf{u}_2\gamma), \dots, p_k(\mathbf{u}_k\gamma)\}$$

induces a finite strongly connected subgraph such that $(\mathbf{u}_i\gamma\theta)^I = (\mathbf{u}_i\theta)^I = \boldsymbol{\xi}_i$. □

### A.5 Proof of Proposition 8

**Proposition 8** *If a rectified formula $F$ in normal form is bounded, then $F$ is bounded w.r.t. any interpretation.*

The proof follows from the following lemma.

**Lemma 16** *Let $F$ be a rectified sentence of a signature $\sigma$ in normal form, and let $I$ be an interpretation of $\sigma$. For any path*

$$\langle p_1(\boldsymbol{\xi}_1^\diamond), p_2(\boldsymbol{\xi}_2^\diamond), \dots, p_k(\boldsymbol{\xi}_k^\diamond), p_{k+1}(\boldsymbol{\xi}_{k+1}^\diamond)\rangle$$

*in the dependency graph of $F$ w.r.t $I$, there exists a path*

$$\langle p_1(\mathbf{u}_1), p_2(\mathbf{u}_2), \dots, p_k(\mathbf{u}_k), p_{k+1}(\mathbf{u}_{k+1})\rangle$$

*in the first-order dependency graph of $F$ with a substitution $\theta$ that maps object variables in $\mathbf{u}_i$ to object names such that $(\mathbf{u}_i\theta)^I = \boldsymbol{\xi}_i$ for all $i$, and $\mathbf{u}_1$ is a list of object variables.*

**Proof**. The proof is similar to the proof of Lemma 14 except that we do not require that $I$ satisfy $\text{CET}_\sigma$. Instead, the existence of a unifier $\gamma$ for $\mathbf{t}'_j\sigma_j^j$ and $\mathbf{t}_{j+1}$ is ensured by the assumption on normal form that $\mathbf{t}_{j+1}$ is a list of variables and the assumption that $\mathbf{t}'_j\sigma_j^j$ contains none of those variables (due to variable renaming).

### A.6 Proof of Proposition 9

**Proposition 9** *If a rectified sentence $F$ in normal form is bounded, then for any interpretation $I$, $I$ is a model of*

$$\{LF_F(Y) \mid Y \textit{ is a finite first-order loop of } F\}$$

*iff $I$ is a model of*

$$\{LF_F(Y) \mid Y \textit{ is a finite loop of } F \textit{ w.r.t. } I\}.$$

The proof follows from Fact 1 and the following lemma.

**Lemma 17** *If a rectified sentence $F$ in normal form is bounded, then for any finite loop $Y'$ of $F$ w.r.t. $I$, there is a finite loop $Y$ of $F$ with a substitution $\theta$ that maps variables in $Y$ to object names such that $Y' = \{p_i(\boldsymbol{\xi}^\diamond) \mid p_i(\mathbf{t}) \in Y,\ (\mathbf{t}\theta)^I = \boldsymbol{\xi}\}$.*

**Proof**. Let $Y'$ be a finite loop of $F$ w.r.t. $I$. Without loss of generality, there is a path

$$\langle p_1(\boldsymbol{\xi}_1^\diamond), p_2(\boldsymbol{\xi}_2^\diamond), \ldots, p_k(\boldsymbol{\xi}_k^\diamond), p_1(\boldsymbol{\xi}_1^\diamond)\rangle$$

($k \geq 1$) in the dependency graph of $F$ w.r.t. $I$ that consists of the vertices in $Y'$. Since $F$ is in normal form, by Lemma 16, there are a path

$$\langle p_1(\mathbf{u}_1), p_2(\mathbf{u}_2), \ldots, p_k(\mathbf{u}_k), p_1(\mathbf{u}_{k+1})\rangle \tag{58}$$

in the first-order dependency graph of $F$, where $\mathbf{u}_1$ consists of object variables only, and a substitution $\theta$ that maps variables in $\mathbf{u}_i$ to object names such that $(\mathbf{u}_i\theta)^I = \boldsymbol{\xi}_i$ for all $1 \leq i \leq k$, and $(\mathbf{u}_{k+1}\theta)^I = \boldsymbol{\xi}_1$. There are two cases to consider.

- *Case 1:* There is a unifier $\gamma$ for $\mathbf{u}_1$ and $\mathbf{u}_{k+1}$ that maps variables in $\mathbf{u}_1$ to terms in $\mathbf{u}_{k+1}$ so that $\mathbf{u}_1\gamma = \mathbf{u}_{k+1}$. It follows that, for any variable $x$ in $\mathbf{u}_{k+1}$ or $\mathbf{u}_1$, $(x\gamma\theta)^I = (x\theta)^I$. Consequently,

  $$\{p_1(\mathbf{u}_1\gamma), p_2(\mathbf{u}_2\gamma), \ldots, p_k(\mathbf{u}_k\gamma)\}$$

  induces a finite strongly connected subgraph such that $(\mathbf{u}_i\gamma\theta)^I = (\mathbf{u}_i\theta)^I = \boldsymbol{\xi}_i$.

- *Case 2:* There is no such unifier $\gamma$.

  Consider another path

  $$\langle p_1(\mathbf{v}_1), p_2(\mathbf{v}_2), \ldots, p_k(\mathbf{v}_k), p_1(\mathbf{v}_{k+1})\rangle$$

  that is obtained similar to (58) except that the variables in the path are disjoint from the variables in (58). Clearly, there is a unifier $\gamma'$ for $\mathbf{u}_{k+1}$ and $\mathbf{v}_1$ that maps the variables $\mathbf{v}_1$ to terms, so that

  $$\langle p_1(\mathbf{u}_1), p_2(\mathbf{u}_2), \ldots, p_k(\mathbf{u}_k), p_1(\mathbf{v}_1\gamma'), p_2(\mathbf{v}_2\gamma'), \ldots, p_k(\mathbf{v}_k\gamma')\rangle$$

  is another path in the first-order dependency graph of $F$. It is clear that using the same construction repeatedly, we can form an infinite path that visits infinitely many vertices in the first-order dependency graph. But this contradicts the assumption that $F$ is bounded.

□

### A.7 Proof of Proposition 11

We will use the following lemma in this section and the next section, which extends Theorem 2 in the work of Chen et al. (2006) that provides a few equivalent conditions for a program to have a finite complete set of loops to a disjunctive program and a sentence.

**Lemma 18** *(Chen et al., 2006, Thm. 2) For any formula $F$ that contains no function constants of positive arity, the following conditions are equivalent:*

*(a) $F$ has a finite complete set of loops.*

(b) *There is a nonnegative integer $N$ such that for every loop $L$ of $F$, the number of variables in $L$ is bounded by $N$.*

(c) *For any loop $L$ of $F$ and any atom $A_1$ and $A_2$ in $L$, the variables occurring in $A_1$ are identical to the variables occurring in $A_2$.*

(d) *For any loop $L$ of $Ground_{\sigma(F)\cup\{c_1,c_2\}}(F)$ where $c_1$, $c_2$ are two new object constants, there are no two atoms $A_1$ and $A_2$ in $L$ such that $A_1$ mentions $c_1$ but $A_2$ does not or $A_1$ mentions $c_2$ but $A_2$ does not.*

**Proposition 11** *For any rectified formula $F$ that contains no function constants of positive arity, $F$ is bounded iff $F$ has a finite complete set of loops.*

**Proof**. *From left to right:* Assume that $F$ is bounded. Then every loop of $F$ is finite. It follows that there exists a nonnegative integer $N$ such that the number of variables in any loop is bounded by $N$. By Lemma 18 (b), $F$ has a finite complete set of loops.

*From right to left:* Assume that $F$ has a finite complete set of loops and, for the sake of contradiction, assume that it is not bounded. Without loss of generality, there is an infinite path

$$\langle p_1(\mathbf{t}_1)\sigma_1, p_2(\mathbf{t}_2)\sigma_2, \dots\rangle \tag{59}$$

in the first-order dependency graph of $F$ that visits infinitely many vertices, where $p_i(\mathbf{t}_i)$ are atoms occurring in $F$ and $\sigma_i$ are substitutions.

Since $F$ is a finite string, it contains finitely many atoms. It follows that there is an atom $p_i(\mathbf{t}_i)$ occurring in $F$ with arbitrarily many substitutions $\sigma$ such that atoms $p_i(\mathbf{t}_i)\sigma$ are contained in (59). Without loss of generality, consider the path

$$\langle p_i(\mathbf{t}_i)\sigma_i, p_{i+1}(\mathbf{t}_{i+1})\sigma_{i+1}, \dots, p_i(\mathbf{t}_i)\sigma_k\rangle$$

that is contained in (59), where $\sigma_k$ and $\sigma_i$ agree on substituting object constants for variables in $\mathbf{t}_i$. Since $\mathbf{t}_i\sigma_i$ and $\mathbf{t}_i\sigma_k$ contain no function constant, there exists a substitution $\sigma_0$ that maps variables in $\mathbf{t}_i\sigma_k$ to terms in $\mathbf{t}_i\sigma_i$ so that $\mathbf{t}_i\sigma_k\sigma_0 = \mathbf{t}_i\sigma_i$. Consequently,

$$\{p_i(\mathbf{x}_i)\sigma_i\sigma_0, p_{i+1}(\mathbf{x}_{i+1})\sigma_{i+1}\sigma_0, \dots, p_i(\mathbf{x}_i)\sigma_k\sigma_0\}$$

is a loop of $F$. Since the length of the path is arbitrarily large, there are arbitrarily many variables occurring in the loop. By Lemma 18 (b), it follows that $F$ has no finite complete set of loops. □

### A.8 Proof of Proposition 10

**Proposition 10** *For any rectified sentence $F$ (allowing function constants of positive arity),*

(a) *checking whether $F$ is bounded is not decidable;*

(b) *checking whether $F$ is atomic-tight is not decidable.*

*If $F$ contains no function constants of positive arity,*

(c) *checking whether $F$ is bounded is decidable;*

(d) *checking whether $F$ is atomic-tight is decidable.*

### A.8.1 Proof of Part (a) and (b)

We show the proof of Part (a) first. The proof repeats, with minor modifications, the argument from the proof of Theorem 26 from the work of Bonatti (2004), which considers the following program $\Pi_{\mathcal{M}}$ to simulate deterministic Turing machines $\mathcal{M}$.

| | |
|---|---|
| $t(s, L, v, [V \mid R], C) \leftarrow t(s', [v' \mid L], V, R, C+1)$ | for all instr. $\langle s, v, v', s', \text{right}\rangle$ |
| $t(s, L, v, [\,], C) \leftarrow t(s', [v' \mid L], b, [\,], C+1)$ | for all instr. $\langle s, v, v', s', \text{right}\rangle$ |
| $t(s, [V \mid L], v, R, C) \leftarrow t(s', L, V, [v' \mid R], C+1)$ | for all instr. $\langle s, v, v', s', \text{left}\rangle$ |
| $t(s, [\,], v, R, C) \leftarrow t(s', [\,], b, [v' \mid R], C+1)$ | for all instr. $\langle s, v, v', s', \text{left}\rangle$ |
| $t(s, L, v, R, C)$ | for all final states s. |

The Halting problem can be reduced to the problem of checking bounded formulas. More precisely, we show that $\Pi_{\mathcal{M}}$ is bounded iff $\mathcal{M}$ terminates from every configuration.

We first establish the following facts:

(i) for every non-terminating computation of $\mathcal{M}$ on input $x$, there is a corresponding infinite path in the first-order dependency graph of $\Pi_{\mathcal{M}}$ that visits infinitely many vertices;

(ii) if there is an infinite path in the first-order dependency graph of $\Pi_{\mathcal{M}}$, then there is an infinite path starting with a legal encoding of an input and corresponds to a non-terminating computation of $\mathcal{M}$.

Fact (i) is immediate from the definition of $\Pi_{\mathcal{M}}$: Note that the step counter (the last argument of $t$) ensures that the dependency graph is acyclic. Then, whenever $\mathcal{M}$ falls into a cycle, the dependency graph contains an infinite acyclic path that visits infinitely many vertices and hence the program is not bounded.

Fact (ii) can be proven as follows. Assume that there is an infinite path in the dependency graph. We observe that the first argument of every vertex in the path must be a legal state and the third argument of every vertex must be a legal tape value. Otherwise, there is no outgoing edge from the vertices in the dependency graph of $\Pi_{\mathcal{M}}$. So only the second, fourth and fifth arguments can contain variables or illegal values which were obtained from substitutions from the variables $L$, $R$, $V$ and $C$. In this case, we can easily find substitutions from these variables or illegal values to legal values and apply them uniformly along the path, so that we obtain another infinite path starting from the vertex that correctly encodes a configuration of $\mathcal{M}$ and thus $\mathcal{M}$ has a corresponding non-terminating computation.

The claim follows immediately from the two facts: if $\mathcal{M}$ does not terminate on some computation, then by (i), $\Pi_{\mathcal{M}}$ is unbounded. If $\Pi_{\mathcal{M}}$ is unbounded, then by (ii), $\mathcal{M}$ does not terminate.

The same proof works for Part (b) as well. This is because the step counter (the last argument of $t$) ensures that the dependency graph is acyclic. Consequently, every infinite path in the dependency graph visits infinitely many vertices, so that $\Pi_{\mathcal{M}}$ is atomic-tight iff $\Pi_{\mathcal{M}}$ is bounded. □

### A.8.2 Proof of Part (c)

In view of the equivalence between (a) and (d) in Lemma 18, checking whether a formula $F$ containing no function constants of positive arity has a finite complete set of loops can be done by examining a finite number of loops from a finite dependency graph, which is decidable. By Proposition 11, it follows that checking whether $F$ is bounded is decidable. □

#### A.8.3 Proof of Part (d)

For any sentence $F$ that has no function constants of positive arity and any finite set $\mathbf{c}$ of object constants, $Ground_{\mathbf{c}}(F)$ is defined recursively. If $F$ is an atomic formula then $Ground_{\mathbf{c}}(F)$ is $F$. The function $Ground_{\mathbf{c}}$ commutes with all propositional connectives; quantifiers turn into finite conjunctions and disjunctions over all object constants occurring in $\mathbf{c}$.

**Lemma 19** *Let* $\mathbf{c}$ *be the set consisting of all object constants occurring in* $F$*, and possibly a new object constant if* $F$ *contains no object constants.* $F$ *has a non-trivial loop iff* $Ground_{\mathbf{c}}(F)$ *has a non-trivial loop.*

In order to check whether $F$ is atomic-tight, we first check whether $F$ is bounded, which is decidable. If $F$ is not bounded, then $F$ is not atomic-tight. Otherwise, in view of Lemma 19, checking whether $F$ is atomic-tight is the same as checking whether $Ground_{\mathbf{c}}(F)$ is atomic-tight. Since $F$ contains no function constants of positive arity, the dependency graph of $Ground_{\mathbf{c}}(F)$ is finite. So it is decidable to check whether the dependency graph of $Ground_{\mathbf{c}}(F)$ has a non-trivial loop. □

### A.9 Proof of Proposition 13

**Lemma 20** *Let* $F$ *be a formula and* $Y$ *a set of atoms. If no predicate constant occurring in* $Y$ *occurs strictly positively in* $F$*, then* $NES_F(Y)$ *is equivalent to* $F$*.*

**Proof**. By induction. □

**Proposition 13** *Let* $\Pi$ *be a program with quantifiers,* $F$ *the FOL-representation of* $\Pi$*, and* $Y$ *a finite set of atoms. Under the assumption* $\Pi$*, formula* $QES_\Pi(Y)$ *is equivalent to* $\neg NES_F(Y)$*. If* $\Pi$ *is a disjunctive program in normal form, then* $QES_\Pi(Y)$ *is also equivalent to* $ES_\Pi(Y)$ *under the assumption* $\Pi$*.*

**Proof**. *Between* $QES_\Pi(Y)$ *and* $\neg NES_F(Y)$*:* $\neg NES_F(Y)$ is

$$\neg \bigwedge_{H \leftarrow G \in \Pi} \forall \mathbf{x}[(G \rightarrow H) \wedge (NES_G(Y) \rightarrow NES_H(Y))]. \tag{60}$$

Under the assumption $F$, formula (60) is equivalent to

$$\bigvee_{H \leftarrow G \in \Pi} \exists \mathbf{x}(NES_G(Y) \wedge \neg NES_H(Y)). \tag{61}$$

In view of Lemma 20, if $H$ does not contain any strictly positive occurrence of a predicate constant that belongs to $Y$, $NES_H(Y)$ is equivalent to $H$. Also, it follows from Lemma 2 and Lemma 8 that $NES_G(Y)$ implies $G$. So $NES_G(Y) \wedge \neg NES_H(Y)$ conflicts the assumption $G \rightarrow H$ when $H$ does not contain any strictly positive occurrence of a predicate constant that belongs to $Y$. As a result, under the assumption $F$, formula (61) is equivalent to the disjunction of

$$\exists \mathbf{x}(NES_G(Y) \wedge \neg NES_H(Y)) \tag{62}$$

for all rules $H \leftarrow G$, where $H$ contains a strictly positive occurrence of a predicate constant that belongs to $Y$. Note that $G$ and $H$ are formulas such that every occurrence of an implication in $G$ and $H$ belongs to a negative formula. By Lemma 4, (62) is equivalent to $QES_\Pi(Y)$.

*Between $QES_\Pi(Y)$ and $ES_\Pi(Y)$:* When $\Pi$ is a disjunctive program, $QES_\Pi(Y)$ is the disjunction of

$$\exists \mathbf{z}\Big(B \wedge N \wedge \bigwedge_{\substack{p(\mathbf{t})\in B \\ p(\mathbf{t}')\in Y}} (\mathbf{t} \neq \mathbf{t}') \wedge \neg( \bigvee_{p(\mathbf{t})\in A} (p(\mathbf{t}) \wedge \bigwedge_{p(\mathbf{t}')\in Y} \mathbf{t} \neq \mathbf{t}'))\Big) \tag{63}$$

over all rules (10) such that $A$ contains a predicate constant that occurs in $Y$, where $\mathbf{z}$ is a list of variables in (10) but not in $Y$. On the other hand, $ES_\Pi(Y)$ is the disjunction of

$$\exists \mathbf{z}'\Big(B\sigma \wedge N\sigma \wedge \bigwedge_{\substack{p(\mathbf{t})\in B\sigma \\ p(\mathbf{t}')\in Y}} (\mathbf{t} \neq \mathbf{t}') \wedge \neg( \bigvee_{p(\mathbf{t})\in A\sigma} (p(\mathbf{t}) \wedge \bigwedge_{p(\mathbf{t}')\in Y} \mathbf{t} \neq \mathbf{t}'))\Big) \tag{64}$$

over all rules (10) such that $A$ contains a predicate constant that occurs in $Y$ and $A\sigma \cap Y \neq \emptyset$, where $\mathbf{z}'$ is a list of variables in $A\sigma \leftarrow B\sigma, N\sigma$ but not in $Y$.

It is clear that (64) implies (63). To prove that (63) implies (64), assume

$$B \wedge N \wedge \bigwedge_{\substack{p(\mathbf{t})\in B \\ p(\mathbf{t}')\in Y}} (\mathbf{t} \neq \mathbf{t}') \wedge \neg( \bigvee_{p(\mathbf{t})\in A} (p(\mathbf{t}) \wedge \bigwedge_{p(\mathbf{t}')\in Y} \mathbf{t} \neq \mathbf{t}')) \tag{65}$$

and consider two cases.
If $\bigwedge_{p(\mathbf{t}')\in Y} \mathbf{t} \neq \mathbf{t}'$ for all $p(\mathbf{t}) \in A$, then (65) is equivalent to

$$B \wedge N \wedge \bigwedge_{\substack{p(\mathbf{t})\in B \\ p(\mathbf{t}')\in Y}} (\mathbf{t} \neq \mathbf{t}') \wedge \neg \bigvee_{p(\mathbf{t})\in A} p(\mathbf{t})$$

which contradicts the assumption $\Pi$.

Otherwise, there exists $p(\mathbf{t}) \in A$ and $p(\mathbf{t}') \in Y$ such that $\mathbf{t} = \mathbf{t}'$. Since $\Pi$ is in normal form, there exists $\sigma$ that maps $\mathbf{t}$ to $\mathbf{t}'$, so that $A\sigma \cap Y \neq \emptyset$. Consequently, (65) is equivalent to

$$B\sigma \wedge N\sigma \wedge \bigwedge_{\substack{p(\mathbf{t})\in B\sigma \\ p(\mathbf{t}')\in Y}} (\mathbf{t} \neq \mathbf{t}') \wedge \neg( \bigvee_{p(\mathbf{t})\in A\sigma} (p(\mathbf{t}) \wedge \bigwedge_{p(\mathbf{t}')\in Y} \mathbf{t} \neq \mathbf{t}')).$$

Thus the claim follows. □

### A.10 Proof of Proposition 16

**Proposition 16** *Let $P$ be an LW-program and let $F$ be the FOL-representation of the set of rules in $P$. The following conditions are equivalent to each other:*

*(a) $I$ is an answer set of $P$ according to Lin and Wang (2008);*

*(b) $I$ is a $P$-interpretation that satisfies* $\mathrm{SM}[F]$*;*

*(c) $I$ is a $P$-interpretation that satisfies $F$ and the loop formulas of $Y$ for all loops $Y$ of $F$ w.r.t. $I$.*

Given a program $\Pi$, $\mathit{Norm}(\Pi)$ is a normal form of $\Pi$ and $\mathit{Ground}(\Pi)$ is a ground program obtained from $\Pi$ as described by Lin and Wang (2008). The proof of Proposition 16 follows from the following lemma. We refer readers to the work of Lin and Wang for the definition of $ES(\cdot,\cdot,\cdot)$ defined there.

**Lemma 21** *For any program $\Pi$ and any set $Y$ of ground atoms, $ES_{\mathit{Norm}(\Pi)}(Y)$ is equivalent to $\bigvee_{p(\mathbf{c})\in Y} ES(p(\mathbf{c}), Y, \mathit{Ground}(\Pi))$.*

**Proof**. By definition, $ES_{\mathit{Norm}(\Pi)}(Y)$ is

$$\bigvee_{\substack{p(\mathbf{x})\leftarrow B,N,\mathbf{x}=\mathbf{t} \text{ is in } \mathit{Norm}(\Pi)\\ \theta: p(\mathbf{x})\theta\in Y}} \exists \mathbf{z}\Big(B\theta \wedge N\theta \wedge \mathbf{x}\theta = \mathbf{t}\theta \wedge \bigwedge_{\substack{q(\mathbf{t})\in B\theta\\ q(\mathbf{t}')\in Y}} (\mathbf{t}\neq\mathbf{t}')\Big), \tag{66}$$

where $\mathbf{x}$ is a list of distinct object variables, $\theta$ is a substitution that maps variables in $\mathbf{x}$ to object constants occurring in $Y$, and $\mathbf{z}$ is the list of all variables that occur in the rule $p(\mathbf{x})\theta \leftarrow B\theta, N\theta, \mathbf{x}\theta = \mathbf{t}\theta$. (66) is equivalent to

$$\bigvee_{\substack{p(\mathbf{t})\leftarrow B,N\in\Pi\\ p(\mathbf{c})\in Y}} \exists \mathbf{z}'\Big(B \wedge N \wedge \mathbf{t} = \mathbf{c} \wedge \bigwedge_{\substack{q(\mathbf{t})\in B\\ q(\mathbf{t}')\in Y}} (\mathbf{t}\neq\mathbf{t}')\Big), \tag{67}$$

where $\mathbf{z}'$ is the list of all variables that occur in the rule $p(\mathbf{t}) \leftarrow B, N$. In turn, (67) is equivalent to

$$\bigvee_{\substack{p(\mathbf{d})\leftarrow B',N'\in \mathit{Ground}(\Pi)\\ p(\mathbf{c})\in Y}} \Big(B' \wedge N' \wedge \mathbf{d} = \mathbf{c} \wedge \bigwedge_{\substack{q(\mathbf{t}_g)\in B'\\ q(\mathbf{t}')\in Y}} (\mathbf{t}_g\neq\mathbf{t}')\Big). \tag{68}$$

Note that when $\mathbf{d}$ does not cover $\mathbf{c}$, there exists $d_i \in \mathbf{d}$ such that $d_i$ mentions only constants and pre-interpreted functions and $d_i$ can not be evaluated to $c_i$ independent of interpretations. In that case, $\mathbf{d} = \mathbf{c}$ is equivalent to $\bot$. Thus (68) is equivalent to

$$\bigvee_{p(\mathbf{c})\in Y} \bigvee_{\substack{p(\mathbf{d})\leftarrow B',N'\in \mathit{Ground}(\Pi)\\ p(\mathbf{d}) \textit{ can cover } p(\mathbf{c})}} \Big(B' \wedge N' \wedge \mathbf{d} = \mathbf{c} \wedge \bigwedge_{\substack{q(\mathbf{t}_g)\in B'\\ q(\mathbf{t}')\in Y}} (\mathbf{t}_g\neq\mathbf{t}')\Big), \tag{69}$$

which is essentially $\bigvee_{p(\mathbf{c})\in Y} ES(p(\mathbf{c}), Y, \mathit{Ground}(\Pi))$. □